\documentclass[useAMS,usenatbib]{mn2e}
\usepackage{amssymb,amsmath}
\usepackage{times}
\usepackage{ctable,longtable}
\usepackage{graphicx}
\usepackage[colorlinks=true, linkcolor=blue, citecolor=blue, urlcolor=blue]{hyperref}
\title[Transition-type dwarfs in the Virgo cluster]{The Herschel Virgo Cluster Survey XIV: transition-type dwarf galaxies in the Virgo cluster}

\author[I. De Looze et al.]{%
Ilse De Looze$^1$, Maarten Baes$^1$, Alessandro Boselli$^2$, Luca Cortese$^{3,4}$, Jacopo Fritz$^1$,  
\newauthor 
Robbie Auld$^5$, George J. Bendo$^6$, Simone Bianchi$^7$, M{\'e}d{\'e}ric Boquien$^2$, Marcel Clemens$^8$, 
\newauthor 
Laure Ciesla$^9$, Jonathan Davies$^5$, Sperello di Serego Alighieri$^7$, Marco Grossi$^{10}$, 
\newauthor 
Anthony Jones$^{11}$, Suzanne C. Madden$^{12}$, Ciro Pappalardo$^{10}$, Daniele Pierini$^{13}$, 
\newauthor 
Matthew~W.~L. Smith$^5$, Joris Verstappen$^1$, Catherine Vlahakis$^{14}$, Stefano Zibetti$^7$  \\
$^1$ Sterrenkundig Observatorium, Universiteit Gent, Krijgslaan 281 S9, B-9000 Gent, Belgium \\
$^2$ Laboratoire d'Astrophysique de Marseille - LAM, Universit\'e d'Aix-Marseille \& CNRS, UMR7326, 38 rue F. Joliot-Curie, 13388 Marseille Cedex 13, France \\
$^3$ European Southern Observatory, Karl Schwarzschild Str. 2, 85748 Garching bei M{\"u}nchen, Germany \\
$^4$ Centre for Astrophysics and Supercomputing, Swinburne University of Technology, PO Box 218, Hawthorn, VIC 3122, Australia \\
$^5$ School of Physics and Astronomy, Cardiff University, The Parade, Cardiff, CF24 3AA, UK \\
$^6$ UK ALMA Regional Centre Node, Jodrell Bank Centre for Astrophysics, School of Physics and Astronomy, \\
University of Manchester, Oxford Road, Manchester M13 9PL, United Kingdom \\
$^7$ INAF-Osservatorio Astrofisico di Arcetri, Largo Enrico Fermi 5, 50125 Firenze, Italy\\
$^8$ INAF-Osservatorio Astronomico di Padova, Vicolo dellÕOsservatorio 5, 35122 Padova, Italy \\
$^{9}$ University of Crete, Department of Physics, Heraklion 71003, Greece \\
$^{10}$ CAAUL, Observat{\'o}rio Astron{\'o}mico de Lisboa, Tapada de Ajuda, 1349-018, Lisboa, Portugal \\
$^{11}$ Institut d'Astrophysique Spatiale (IAS), UMR 8617, CNRS/Universit{\'e} Paris-Sud, 91405 Orsay, France\\
$^{12}$ Laboratoire AIM, CEA/DSM - CNRS - Universit{\'e} Paris Diderot, Irfu/Service d'Astrophysique, CEA Saclay, 91191 Gif-sur-Yvette, France \\
$^{13}$ Visiting astronomer at the Max-Planck-Institut fŸr extraterrestrische Physik (MPE), Giessenbachstrasse, 85748 Garching, Germany \\
$^{14}$ Joint ALMA Observatory / European Southern Observatory, Alonso de Cordova 3107, Vitacura, Santiago, Chile}

\begin{document}

\date{Accepted 2013 August 27.  Received 2013 August 18; in original form 2013 May 13}

\pagerange{\pageref{firstpage}--\pageref{lastpage}} \pubyear{2013}

\maketitle

\label{firstpage}

\begin{abstract}
We use dust scaling relations to investigate the hypothesis that Virgo cluster transition-type dwarfs are infalling star-forming field galaxies, which is argued based on their optical features (e.g. disks, spiral arms, bars) and kinematic properties similar to late-type galaxies. After their infall, environmental effects gradually transform them into early-type galaxies through the removal of their interstellar medium and quenching of all star formation activity. 

In this paper, we aim to verify whether this hypothesis holds using far-infrared diagnostics based on \textit{Herschel} observations of the Virgo cluster taken as part of the \textit{Herschel} Virgo Cluster Survey (HeViCS). We select transition-type objects in the nearest cluster, Virgo, based on spectral diagnostics indicative for their residual or ongoing star formation. We detect dust ($M_{\text{d}} \sim 10^{5-6}~M_{\odot}$) in 36$\%$ of the transition-type dwarfs located on the high end of the stellar mass distribution. This suggests that the dust reservoirs present in non-detections fall just below the \textit{Herschel} detection limit ($\lesssim 1.1\times10^5~M_{\odot}$). 

Dust scaling relations support the hypothesis of a transformation between infalling late-type galaxies to quiescent low-mass spheroids governed by environmental effects, with dust-to-stellar mass fractions for transition-type dwarfs in between values characteristic for late-type objects and the lower dust fractions observed in early-type galaxies.
Several transition-type dwarfs demonstrate blue central cores, hinting at the radially outside-in removal of gas and quenching of star formation activity. 
The fact that dust is also confined to the inner regions suggests that metals are stripped in the outer regions along with the gas. 
In the scenario of most dust being stripped from the galaxy along with the gas, we argue that the ejected metals by transition-type dwarfs significantly contribute to the enrichment of the intra-cluster medium over the lifetime of the Virgo cluster. The accretion of gas through tidal interactions and re-ignition of star formation in the centres of transition-type dwarfs could provide an alternative explanation for the observed dust scaling relations and blue central cores.
\end{abstract}


\section{Introduction}

Early-type dwarf galaxies (dEs) are the dominant galaxy population in clusters and originally thought to constitute a homogeneous population of dwarfs devoid of a cool interstellar medium (ISM) and dominated by an old stellar population. However, the detection of spiral arms and/or bars \citep{1991A&A...252...27B,2000A&A...358..845J,2002A&A...391..823B,2003A&A...400..119D,2003AJ....125.2936G,2006AJ....132..497L,2010ApJ...711L..61M,2012ApJ...745L..24J}, the presence of blue central cores \citep{1984A&A...139L...9V,2006AJ....131..806G,2006AJ....132.2432L}, the UV emission from residual star formation \citep{2005ApJ...629L..29B} and evidence for bulge and disk components \citep{2005AJ....130..475A,2012ApJS..198....2K} in a subset of the dwarf elliptical galaxy population clearly show that {not all early-type dwarf galaxies seem to fit this picture of structural homogeneity.} Also the dependence of flattening on the eccentricity of their orbits through the cluster \citep{2009ApJ...706L.124L} seems to indicate that the environment plays an important role in the formation of at least a significant subpopulation of early-type dwarf galaxies.  The rotational support of their stellar kinematics \citep{2002A&A...384..371S, 2002MNRAS.332L..59P,2003A&A...400..119D,2003AJ....126.1794G,2004AJ....128..121V,2009ApJ...707L..17T,2011A&A...526A.114T,2012A&A...548A..78T,2013MNRAS.428.2980R} and the identification of features characteristic of late-type spiral or irregular galaxies in early-type dwarfs support the idea that at least a fraction of the early-type dwarf galaxy population is formed through the transformation of infalling late-type field galaxies by environmental mechanisms \citep{2008ApJ...674..742B,2008A&A...489.1015B}. The detection of significant amounts of gas \citep{2002ApJ...573L...5C,2004AJ....128..121V,2007A&A...474..851D} and dust (\citealt{2010A&A...518L..54D}, \citealt{2013A&A...552A...8D}) in some dwarf ellipticals (dEs) might also reflect their descent of a late-type precursor population. The signatures for the infall of field galaxies from their dynamics \citep{1993A&AS...98..275B,2001ApJ...548L.139D,2001ApJ...559..791C,2008ApJ...674..742B} demonstrate that this formation through environmental effects in clusters only came about during the last few Gyr, with a precursor population of infalling field objects being morphologically different from early-type dwarf galaxies. The similar slope of the luminosity function observed for field \citep{2005ApJ...631..208B} and Virgo cluster galaxies \citep{1985AJ.....90.1759S} furthermore supports this scenario of transforming infalling field galaxies.

{The transformation of infalling field galaxies through environmental effects is, however, not the only possible evolutionary hypothesis for the subsample of early-type dwarf galaxies containing a detectable ISM reservoir and showing evidence for ongoing or recent star formation activity.Rather than descending from the late-type galaxy population, some early-type dwarf galaxies might have re-accreted material through tidal interactions with the cluster medium or a neighboring galaxy, which could also give rise to the re-ignition of star formation. Some observations indeed seem to contradict a transformation from late-type spirals or blue compact dwarfs in the field to early-type dwarf galaxies in clusters.} The bright nuclei observed in nucleated dwarf elliptical galaxies \citep{1985AJ.....90.1681B,1994A&ARv...6...67F,2006ApJS..165...57C,2006ApJ...644L..21F} are incompatible with {the absence of a compact star cluster in the centres of Magellanic irregulars}. Limitations in the resolution of the data are, however, often considered to be at the cause of this incompatibility. Also the mass specific frequency of globular clusters ($T_{\text{N}}$ = $N_{\text{gc}}$/($M_{\star} / 10^9~M_{\odot}$), i.e. the number of globular clusters normalized by stellar mass) differs among dwarf elliptical galaxies and star-forming dwarfs \citep{1998ApJ...508L.133M,2006AJ....132.2333S,2007ApJ...670.1074M,2008ApJ...681..197P,2012MNRAS.424.2614S}, with only low mass early-type galaxies (ETGs, $M_{\star}\leq 2\times10^8~M_{\odot}$) harbouring globular cluster systems (GCS) with a mass specific frequency and spatial distribution resembling gas-rich irregular or late-type dwarfs \citep{2008ApJ...674..742B,2012MNRAS.424.2614S}. {Higher mass ETGs have properties (i.e. high mass specific frequencies of GCSs and a concentrated spatial distribution within Virgo) which seems incompatible with a recent, environmentally driven evolution \citep{2012MNRAS.424.2614S}.}

In this paper, we focus on low-luminosity, early-type dwarf galaxies located in the nearest cluster, Virgo, showing evidence of ongoing or residual star formation activity. {Owing to their shallow potential wells, external interactions (ram-pressure stripping, galaxy harassment, tidal interactions) and/or internal effects (supernova explosions) can more easily overcome the gravitational binding of interstellar medium (ISM) material in low-mass objects compared to giant spirals (e.g. \citealt{2000ApJ...538..559M,2003MNRAS.345.1329M}), resulting in a significant atomic gas reservoir being expelled from the galaxy \citep{2004MNRAS.353.1293M,2004AJ....128..121V,2005A&A...433..875R,2008ApJ...674..742B,2008A&A...489.1015B,2008MNRAS.389.1111V,2012A&A...543A.112A}.} The removal of most of the ISM content is capable of ceasing star formation activity on relatively short timescales, transforming them into quiescent dwarf elliptical galaxies \citep{2010A&A...517A..73G}. During the process of transformation, transition-type objects show properties in between late-type spiral and quiescent early-type galaxies (see \citealt{2008ApJ...674..742B} and \citealt{2013MNRAS.428.2949K} for a more detailed description on their identification and definition). For those low-luminosity galaxies, ram-pressure stripping \citep{1972ApJ...176....1G} has been argued to be the dominant transformation mechanism \citep{2008ApJ...674..742B}. A possible intervention of gravitational interactions (e.g. galaxy harassment, \citealt{2005MNRAS.364..607M}; tidal stirring, \citealt{2001ApJ...547L.123M,2001ApJ...559..754M,2006MNRAS.369.1021M}; starvation, \citealt{1980ApJ...237..692L}) in their formation can, however, not be ruled out based on the predictions of models for different structural scaling relations \citep{2008A&A...489.1015B}. An alternative scenario suggests the re-accretion of gas, which delivers a fresh gas supply and could re-ignite the star formation activity \citep{2012AJ....144...87H,2013ApJ...770L..26D}. Similar re-accretion scenarios are considered as a plausible origin for the observed  dust reservoirs in more massive, early-type galaxies \citep{2012ApJ...748..123S,2013A&A...552A...8D}.

A detailed overview of all environmental effects working on galaxies in clusters is provided in \citet{2006PASP..118..517B}, along with mathematical prescriptions to infer the efficiency of each mechanism throughout the cluster environment. With this diversity of environmental effects being characterized by different timescales and efficiencies, the formation of a single galaxy is most likely influenced by a combination of several processes. Since the outcome of these interactions on a galaxy's morphology and kinematics is typically only observable within a short time range, it is often a strenuous task to identify the dominant transformation process at work, requiring a multi-wavelength dataset of observations, preferentially complemented with knowledge on the stellar kinematics (e.g. \citealt{2009ApJ...707L..17T,2011A&A...526A.114T}). 

With the advent of the \textit{Herschel} Space Observatory \citep{2010A&A...518L...1P}, this large inventory of data can be extended to far-infrared and submillimeter wavelengths. In this work, we benefit from this extended wavelength coverage to probe the far-infrared (FIR) properties of transition-type dwarf galaxies. More specifically, \textit{Herschel} observations covering the central regions of the Virgo cluster (\textit{Herschel} Virgo Cluster Survey, HeViCS, \citealt{2010A&A...518L..48D})  allow us to study the FIR properties of the transition-type population for the very first time. We furthermore investigate whether their dust scaling relations support the hypothesis of them being the transformation products of infalling star-forming dwarfs and try to identify the possible progenitor populations. Observational details are provided in Section \ref{Herscheldata.sec}. The selection of transition-type objects in Virgo is outlined in Section \ref{Selection.sec}. The analysis of the FIR data, including the identification of FIR detections and the dust scaling relations are discussed in Section \ref{FIR.sec}. Section \ref{Description.sec} provides an overview of the general properties of the selected sample, related to their optical morphology, H{\sc{i}} content and cluster location. Section \ref{Dustdiscussion.sec} discusses the environmental processes governing their evolution and provides estimates for the metal enrichment of the intra-cluster medium through metals expelled from transition-type galaxies. Finally, Section \ref{Conclusions.sec} sums up our results. Similar to the abbreviations for other morphological classifications, we will denote transition-type dwarfs as ``TTD'' throughout this paper.

\section{Herschel observations and photometry}
\label{Herscheldata.sec}

The \textit{Herschel} Virgo Cluster Survey (HeViCS, \citealt{2010A&A...518L..48D}) covered 84 sq. degrees of the Virgo cluster with the PACS (100, 160 $\mu$m) and SPIRE (250, 350, 500 $\mu$m) photometers onboard \textit{Herschel} in parallel fast-scan (60$\arcsec$/s) mode. Four overlapping tiles of 4$\times$4 sq. degrees were observed, extending from the central regions around M87 to the NW, W and S cloud structures. Each one of these tiles was covered four times in two orthogonal scan directions, to reduce the $1$/$f$ noise in the final maps.

PACS photometry maps at 100 and 160\,$\mu$m were obtained through a naive projection with photProject in HIPE (v7.3.0). High-pass filtering techniques with filter lengths of 10 and 20 frames at 100 and 160\,$\mu$m, respectively, were applied to reduce the $1$/$f$ noise. Bright sources were masked during this procedure to prevent any ``overshooting'' (i.e. the removal of flux from the brightest source regions). Final PACS maps were constructed with pixel sizes of 2$\arcsec$ and 3$\arcsec$ at 100 and 160\,$\mu$m, which have average beam sizes (i.e. full-width half-maximum, FWHM) of 9.4$\arcsec$ and 13.4$\arcsec$, respectively\footnote{in agreement with http://herschel.esac.esa.int/Docs/PACS/pdf/pacs$\_$om.pdf}. Noise measurements in the final PACS maps combining all 8 scans indicate that background uncertainties are dominated by instrumental noise with values of 1.9 and 1.2 mJy pixel$^{-1}$ in the 160 and 100\,$\mu$m bands, respectively, and decrease down to 1.3 and 0.8 mJy pixel$^{-1}$ in the overlapping map areas of two tiles \citep{2013MNRAS.428.1880A}. Calibration uncertainties for PACS {have been shown to consist} of uncorrelated uncertainties of 3$\%$ and 4$\%$ at 100 and 160\,$\mu$m, respectively, and a correlated noise factor of 2.2$\%$. With the previous estimates of the calibration uncertainty being derived for point sources, which were observed, reduced and analyzed differently than our HeViCS PACS observations, we safely assume a conservative uncertainty of 12$\%$ for the PACS calibration \citep{2013MNRAS.428.1880A}.

The SPIRE timeline data were processed using a variant of the standard data processing pipeline \citep{2010A&A...518L...3G,2010SPIE.7731E.101D} in which the BriGAdE method (Smith et al. 2013, in prep.) was used in place of the standard temperature drift removal module. Final SPIRE maps were constructed with pixel sizes of 6$\arcsec$, 8$\arcsec$ and 12$\arcsec$ at 250, 350 and 500\,$\mu$m, respectively. The FWHM of the beam corresponds to 18.2$\arcsec$, 25.4$\arcsec$ and 36.0$\arcsec$ in the 250, 350 and 500\,$\mu$m maps, respectively, and beam areas have sizes of 423$\arcsec^2$, 751$\arcsec^2$ and 1587$\arcsec^2$, respectively\footnote{in agreement with http://herschel.esac.esa.int/Docs/SPIRE/pdf/spire$\_$om.pdf}. {We are aware of the most recent update regarding the SPIRE Photometer Beam Profiles, estimating beam areas of 465$\arcsec^{2}$, 822$\arcsec^{2}$ and 1768$\arcsec^{2}$.\footnote{http://herschel.esac.esa.int/twiki/bin/view/Public/SpirePhotometer\-Beam\-ProfileAnalysis} For ease of comparison with quoted values in the literature (e.g. \citealt{2012A&A...540A..52C}), we decided not to recalculate the flux measurements based on the new SPIRE beam sizes nor redo the SED fitting procedures as quoted in \citet{2013MNRAS.428.1880A} and \citet{2013A&A...552A...8D}. Accounting for the recent values for the beam areas would not significantly alter the estimated dust masses and temperatures. For upper dust mass limits, the dust mass would be 9$\%$ lower relying on the new beam areas, which corresponds to a shift of 0.04 dex for the average 5$\sigma$ dust mass detection limit $M_{\text{d}}\sim1.1\times10^5~M_{\odot}$ in the \textit{Herschel} 250\,$\mu$m map (see Section \ref{modbb.sec}). Redoing the SED fitting procedure for FIR detected TTDs with the latest beam measurements shows that the resulting dust temperatures will be at most 0.5 K higher and the dust masses at most 20$\%$ (or 0.08 dex) lower and, thus, only change within the error bars of the fitting procedures.} Typical noise levels attained in the final SPIRE maps combining all 8 scans range from 6.6, 7.3 to 8.0 mJy beam$^{-1}$ at 250, 350 and 500\,$\mu$m, respectively, among which half of the uncertainties are attributed to instrumental noise and the remainder is due to confusion noise. Calibration uncertainties for SPIRE are assumed to be around 7$\%$ in each band, including the 5$\%$ correlated error from the assumed models used for Neptune and a random uncertainty of 2$\%$ accounting for the repetitive measurements of Neptune (see SPIRE Observers' Manual: http://herschel.esac.esa.int/Docs/SPIRE/pdf/spire$\_$om.pdf). A more recent study estimate the uncertainties on the flux calibration to be even lower (1.5$\%$, \citealt{Bendo2013}). More details about the data reduction and photometry of \textit{Herschel} observations from the \textit{Herschel} Virgo Cluster Survey (HeViCS) are outlined in \citet{2013MNRAS.428.1880A}.

\section{Sample selection}
\label{Selection.sec}

We gather all Virgo cluster galaxies from the Virgo Cluster Catolog (VCC) which are classified as early-type objects based on the updated morphological classification provided by the Goldmine database \citep{2003A&A...400..451G}. Among the early-type objects, we consider morphological classifications ranging from elliptical (dE, E) {to lenticular} (dS0, S0, S0a) objects, with the inclusion of mixed galaxy types (dE/dS0, E/S0, S0/Sa). We furthermore exclude all background sources with distances $>$ 32 Mpc. Distances are retrieved from the Goldmine database, based on results reported in \citet{1999MNRAS.304..595G}. Since low-luminosity galaxies are easily susceptible to environmental effects in clusters due to their low masses, we restrict the sample to objects with $H$ band luminosities $L_{\text{H}} < 10^{9.6}~L_{\odot}$ (similar to \citealt{2008ApJ...674..742B}), with $L_{\text{H}}$ {tracing the dynamical mass of galaxies within their optical extent \citep{1996A&A...312..397G}\footnote{The $H$ band luminosity has been found to correlate with the dynamical mass within the optical radius of galaxies (e.g. \citealt{1996A&A...312..397G}). This correlation will likely not hold beyond the optical extent due to the dominant mass contribution from dark matter at larger radii.}.} $H$ band photometry is obtained from the Goldmine database, gathering data from several observing campaigns \citep{1996A&AS..120..489G,1996A&AS..120..521G,2000A&AS..142...65G,2001A&A...372...29G}. Due to the incomplete $H$ band coverage, the sample was constrained in practice by limiting the photographic magnitude to $m_{\text{pg}}$ $\ge$ 13.5 (which corresponds to $M_{\text{B}}$ $\gtrsim$ -18 mag and roughly to $L_{\text{H}} < 10^{9.6}~L_{\odot}$ at a distance of 17 Mpc). To guarantee the optical completeness, we delimit our sample to galaxies with photographic magnitudes $m_{\text{pg}}$ $\le$ 18.0, which is also the completeness limit of the VCC catalogue from which our sample is selected. Given our interest in the dust properties of TTDs, we select only objects covered by \textit{Herschel}, resulting in an optically complete sample of 395 early-type Virgo members with \textit{Herschel} observations. 

\begin{figure} 
\centering \includegraphics[width=0.45\textwidth]{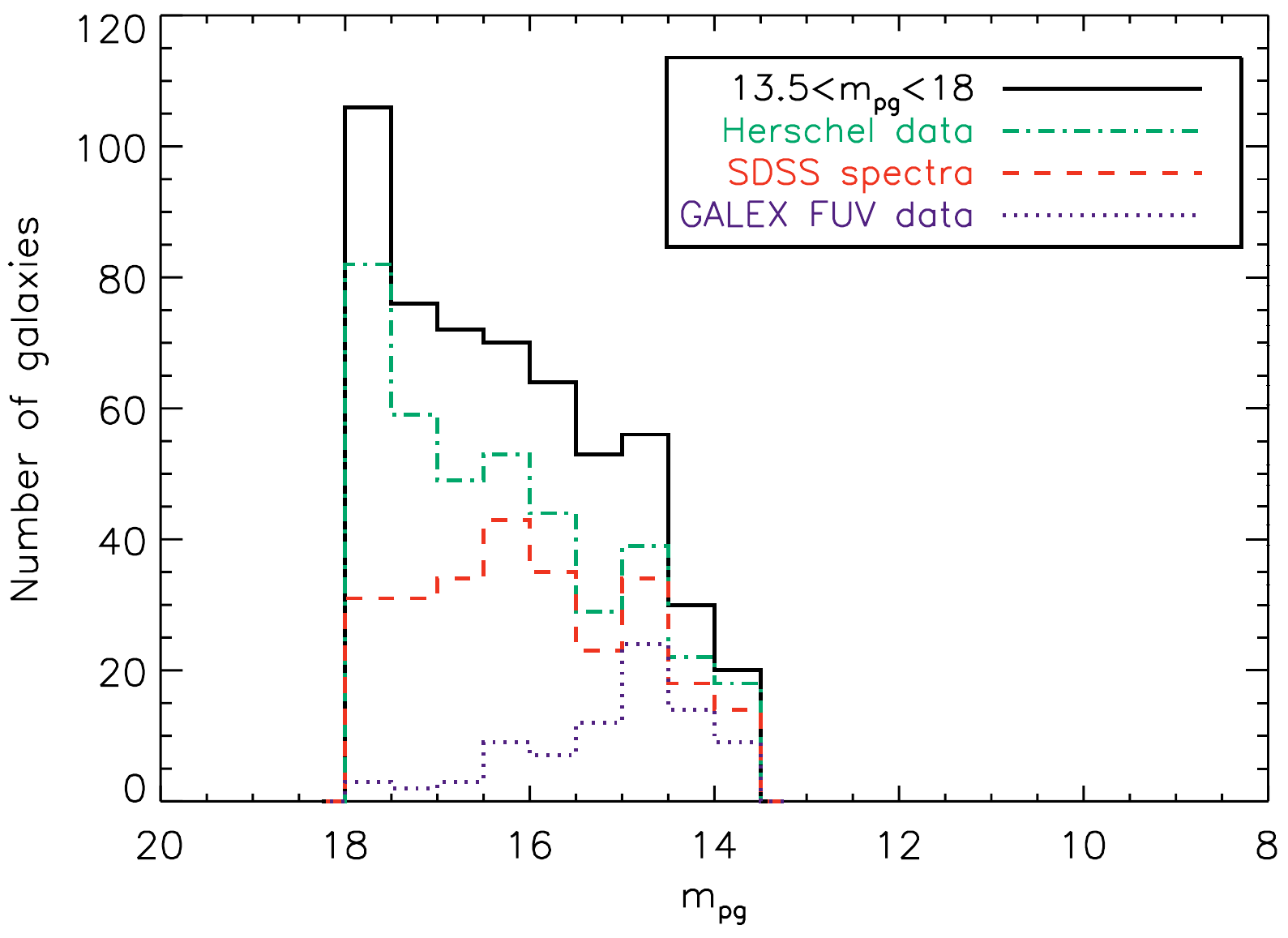}  \\ 
\caption{Histogram representing the galaxy selection procedure. The initial optically complete sample of low-luminosity early-type galaxies (black solid line) is restricted to the objects observed by \textit{Herschel} (green dashed-dotted line). The availability of SDSS spectra (red dashed line) and GALEX $FUV$ data (purple dotted line) for the latter sample is indicated as well.}   \label{plot_histo}
\end{figure}

We then select TTD galaxies in the Virgo cluster based on the identification of residual or ongoing star formation, similar to the selection procedure in \citet{2008ApJ...674..742B}. Considering the quiescent nature of genuine ellipticals, our selection procedure will target objects with an apparent early-type morphology which at the same time harbour the left-over star-forming characteristics of late-type galaxies. Spectral diagnostics obtained from the Sloan Digital Sky Survey (SDSS) are used to probe residual star formation activity in the central regions of galaxies. With spectroscopic data retrievable from the SDSS DR7 archive for 261 objects \citep{2009ApJS..182..543A}, the star forming conditions can be probed for 66$\%$ of the sample galaxies (see red, dashed line in Figure \ref{plot_histo}). 

The selection of transition-type objects from SDSS spectroscopy data is based on the equivalent width of Balmer lines: star-forming galaxies are characterized by strong H$\alpha$ emission (H$\alpha$ EW$_{\text{em}}$~$>$~2\,\AA, \citealt{2008ApJ...674..742B}), whereas post-starburst objects can be distinguished from their H$\beta$ (H$\beta$ EW$_{\text{abs}}$~$>$~2.8\,\AA, \citealt{2008ApJ...674..742B}) or H$\delta$ (H$\delta$ EW$_{\text{abs}}$~$>$~4.0\,\AA, e.g. \citealt{1999ApJ...518..576P,2003PASJ...55..771G}) absorption lines. SDSS spectra are analyzed with a spectral fitting code \citep{2007A&A...470..137F} to determine H$\alpha$, H$\beta$ and H$\delta$ equivalent widths and afterwards checked by eye to verify the spectral line measurement. During the spectral fitting, the main features of the observed spectrum, namely the flux in various continuum bands and the equivalent width of the most significant lines, are reproduced by summing the theoretical spectra of Single Stellar Populations (SSPs) of different ages, ranging from $\sim$ 2 $\times$ 10$^6$ yr to $\sim$ 14 Gyr (12 in total). The latter are built with the Padova evolutionary models \citep{1994A&AS..106..275B}, and use the \citet{1984ApJS...56..257J} spectral libraries. A \citet{1955ApJ...121..161S} IMF with stellar masses between 0.15-120 $M_\odot$ is assumed. The theoretical spectra also include emission lines (Hydrogen, Oxygen, Nitrogen and Sulfur) which are calculated with Cloudy. These affect the optical spectra of SSPs younger than 2 $\times$ 10$^7$ years. A homogenous value of the metallicity is assumed within a given set of SSPs, but globally 3 different sets are tested: Z=0.004, Z=0.02 (solar) and Z=0.05. Before the SSP spectra are summed together, each one is weighted by a certain stellar mass and, furthermore, a given extinction value is applied. The Galactic extinction curve is adopted and dust is assumed to be distributed in a uniform slab in front of the stars. Allowing the extinction to vary as a function of stellar age, is consistent with the selective extinction picture with younger stars being most affected by dust extinction, still embedded in the molecular clouds where they were born. Stellar mass and the amount of extinction are the free parameters of the fitting problem for each of the 12 theoretical spectra. Uncertainties on the equivalent width measurements are calculated in a conservative way, including the noise introduced by the continuum subtraction and the actual equivalent width measurement of the line. Excluding unreliable identifications based on noisy spectra {(i.e. equivalent width measurements with signal-to-noise levels per {\AA}ngstr{\"o}m lower than 3$\sigma$)}, we identify 33 objects with recent or ongoing star formation activity, i.e. a selection rate of 13$\%$. Based on this spectral classification, 15 and 18 galaxies were classified as star-forming and post-starburst objects based on star formation diagnostics in their centres, respectively.

\begin{figure} 
\centering \includegraphics[width=0.45\textwidth]{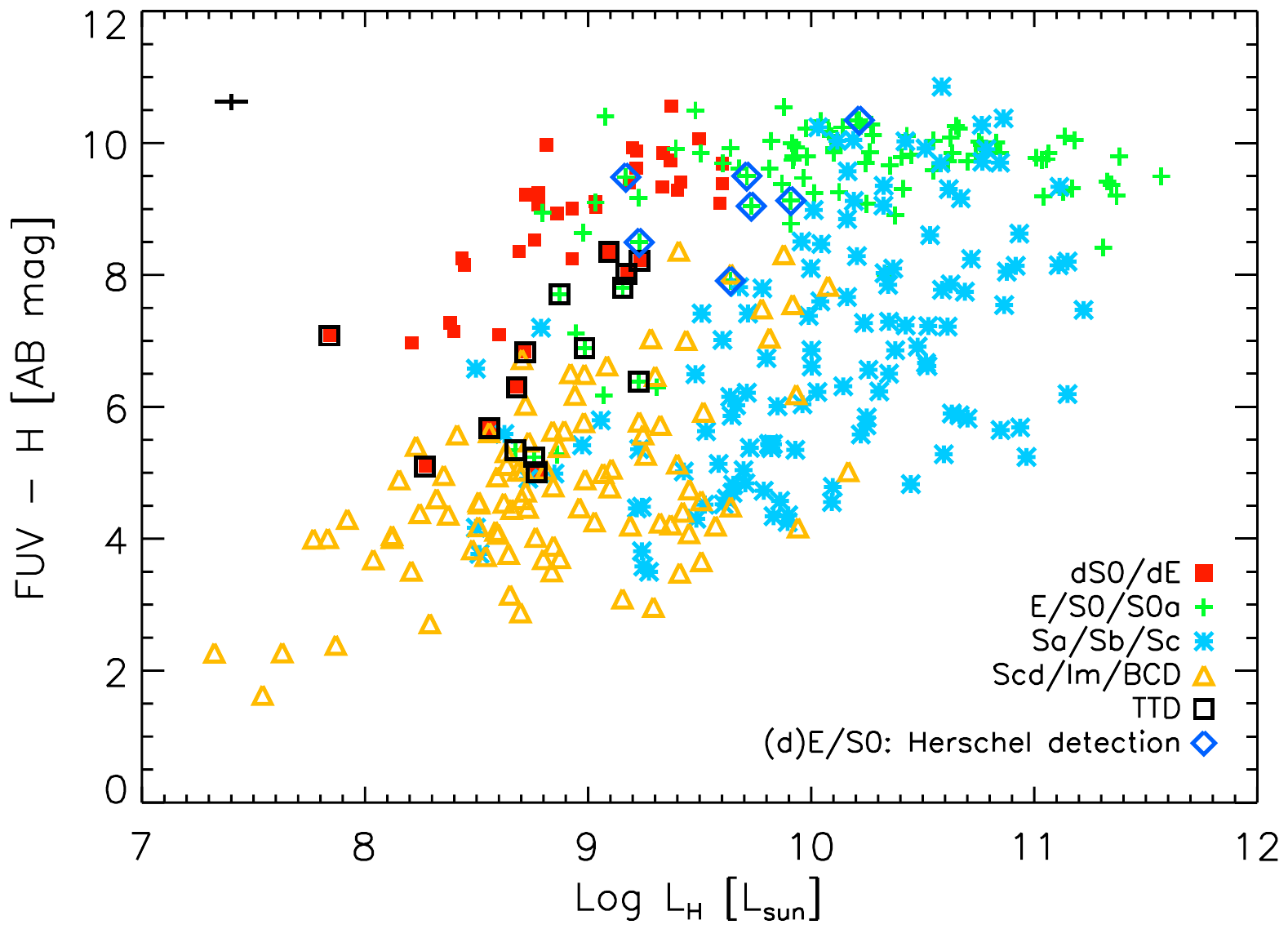}  \\ 
\caption{Plot of the $H$ band luminosity versus the $FUV$-$H$ colours of galaxies. The legend clarifies the symbols for the different morphological populations. The average uncertainty level for data points is indicated in the top left corner.}  \label{plot_UV_H}
\end{figure}

Among the galaxies lacking SDSS spectroscopy data but with GALEX $FUV$ data available\footnote{Due to the recent break down of the FUV filter onboard the GALEX space observatory, the number of objects with FUV photometry is limited (see purple, dotted line in Figure \ref{plot_histo}).}, we classify another three galaxies with obvious blue colours as TTD (see Figure \ref{plot_UV_H}): VCC\,1501 (FUV-H $\sim$ 5.7), VCC\,1512 (FUV-H $\sim$ 7.1) and VCC\,1715 (FUV-H $\sim$ 5.1). $FUV$ and $NUV$ magnitudes are retrieved from the new GALEX catalog of VCC galaxies (Voyer et al. in prep.), based on data of the GALEX Ultraviolet Virgo Cluster Survey (GUVICS, \citealt{2011A&A...528A.107B}). The interpretation of the FUV emission from galaxies might be complicated by the UV upturn (see also \citealt{2005ApJ...629L..29B}), relating the FUV emission to the evolved stellar population rather than to star formation in low-metallicity giant ellipticals \citep{2008ASPC..392....3Y}. With our sample consisting of low-mass objects, their $FUV-H$ colours should not be affected by any UV upturn. Relying on the $FUV-H$ range (5.5 $\lesssim$ $FUV-H$ $\lesssim$ 8.0) covered by TTDs (i.e. $\mu$ $\pm$ $\sigma$, where $\mu$ and $\sigma$ represent the mean and dispersion in the $FUV-H$ distribution), we are confident that the selected galaxies have colours characteristic of their ongoing/recent star formation activity.

In summary, we select a sample of 33 transition-type objects based on their spectral properties. Another three galaxies with blue colours are added to this sample. Due to the lack of SDSS spectrophotometric and GALEX $FUV$ data, we likely fail to select about one-third of the transition-type population observed by \textit{Herschel}. Also the restriction of our sample to TTDs in the HeViCS fields, covering the central 84 sq. degrees of the Virgo cluster, might bias our selection with many TTDs also present in the outer Virgo regions \citep{2013MNRAS.428.2949K}. For some galaxies with available SDSS spectra and GALEX photometry, their blue $FUV-H$ colours suggest recent star formation activity (see Figure \ref{plot_UV_H}). We, however, do not select these galaxies as TTDs since the spectral diagnostics of their SDSS spectra not insisted on their selection. Rather than relaxing our selection criteria and risking any false identifications, we restrict our study to the former 36 selected TTDs.

Table \ref{transition} summarizes their morphological classification, distance, position within sub-clusters or clouds in Virgo, $FUV-H$ colour and spectral diagnostics (H$\alpha$, H$\beta$, H$\delta$ EW). Equivalent line width measurements and $FUV$-$H$ colours resulting in the classification of galaxies as TTDs are indicated in boldface.

\begin{table*}
\caption{Transition-type galaxies in Virgo with their VCC number, morphological classification, distance D, position within sub-clusters or clouds in Virgo, $FUV-H$ colours, H$\alpha$, H$\beta$ and H$\delta$ equivalent width. Equivalent width measurements with a positive and negative sign indicate lines in absorption and emission, respectively. } 
\label{transition}
\begin{center}
\begin{tabular}{@{}|l|c|c|c|c|c|c|c|}
\hline
VCC & type & $D$ & Pos & $FUV-H$\footnotemark[1]\footnotemark[2] & H$\alpha$ EW\footnotemark[1] & H$\beta$ EW\footnotemark[1] & H$\delta$ EW\footnotemark[1] \\
& & [Mpc] & & [AB mag] & [\AA] & [\AA] & [\AA] \\
\hline
209 & dS0 & 17 & N & - &  {-15.12 $\pm$ 1.94} & -1.52 $\pm$ 1.06 & 1.65 $\pm$ 0.65 \\
278 & dS0 & 23 & B & 8.22 & {-9.58 $\pm$ 1.55} & 2.86 $\pm$ 1.85 & 2.51 $\pm$ 1.26 \\
327 & S0 & 32 & W & 6.38 & {-19.75 $\pm$ 2.22} & 2.16 $\pm$ 1.06 & 4.50 $\pm$ 1.21 \\
450 & S0 & 23 & B & 6.89 & {-64.83 $\pm$ 4.03} & -7.26 $\pm$ 2.40 & 1.84 $\pm$ 0.73 \\
571 & S0 & 23 & B & 7.81 & {-17.06 $\pm$ 2.07} & 1.93 $\pm$ 1.03 & 3.13 $\pm$ 1.14 \\
710 & dS0 & 17 & S & 5.01 & {-33.78 $\pm$ 2.91} & -4.10 $\pm$ 1.52 & 2.68 $\pm$ 0.97 \\
781 & dS0 & 17 & A & 8.35 & / & 1.45 $\pm$ 0.66 & {4.38 $\pm$ 1.41} \\
788 & dE & 17 & A & - & 2.54 $\pm$ 0.85 & 0.52 $\pm$ 1.91 & {5.06 $\pm$ 1.57} \\
951 & dE/dS0 & 17 & A & 8.02 & 1.76 $\pm$ 0.79 & 3.18 $\pm$ 1.02 & {5.82 $\pm$ 1.62} \\
1175 & E-E/S0 & 23 & B & 5.34 & {-171.94} $\pm$ 6.56 & -35.43 $\pm$ 3.45 & / \\
1512 & dS0 & 17 & A & {7.09} & - & - & - \\
1684 & dS0 & 17 & A & 6.83 & {-4.65 $\pm$ 1.08} & 1.87 $\pm$ 1.28 & 3.13 $\pm$ 1.19 \\
1715 & dE & 17 & S & {5.10} & - & - & - \\
\hline
181 & dE & 17 & N & - & {-86.25 $\pm$ 4.64} & -15.11 $\pm$ 2.23 & 0.38 $\pm$ 0.64 \\ 
216 & dE & 32 & W & - &  0.50 $\pm$ 0.47 & {3.74 $\pm$ 1.42} & {4.39 $\pm$ 1.36} \\
282 & dE & 32 & W & - &  {-4.05 $\pm$ 1.0}1 & 1.20$\pm$ 0.67 & / \\
304 & dE & 17 & N & - &  {-19.32 $\pm$ 2.20} & -1.14 $\pm$ 1.41 & / \\
409 & dE & 23 & B & - &  {-11.40 $\pm$ 1.69} & / & 7.80 $\pm$ 5.09 \\
525 & dE & 23 & B & -  & {-10.23 $\pm$ 1.60} & 0.91 $\pm$ 1.07 & / \\
611 & dE & 23 & B & - &  {-18.83 $\pm$2.17} & -4.39 $\pm$ 1.26 & {6.98 $\pm$ 2.96} \\
855 & dE & 23 & B & - &  2.48 $\pm$ 0.93 & -3.15 $\pm$ 3.69 & {10.72 $\pm$ 1.96} \\
953 & dE & 17 & A & - &  2.67 $\pm$ 0.93 & {3.18 $\pm$ 0.93} & {4.63 $\pm$ 1.27} \\
1005 & dE & 17 & A & - &  3.64 $\pm$ 1.11 & 2.53 $\pm$ 3.02 & {4.97 $\pm$ 1.55} \\
1039 & dE & 17 & A & - &  / & {8.31 $\pm$ 2.60} & / \\
1078 & dE & 23 & B & 6.30 &  {-3.72 $\pm$ 0.96} & 0.10 $\pm$ 1.14 & 3.89 $\pm$ 1.06 \\
1095 & dE & 17 & S & - &  4.60 $\pm$ 1.52 & -3.03 $\pm$ 2.36 & {7.68 $\pm$ 2.60} \\
1129 & dE & 17 & A & - &  4.73 $\pm$ 1.23 & {7.84 $\pm$ 1.45} & 7.62$\pm$ 7.16 \\
1222 & dE/dS0 & 17 & A & - & / & 1.83 $\pm$ 1.40 & {4.03 $\pm$ 1.15} \\
1314 & dE & 17 & A & - &  / & 6.41 $\pm$ 7.37 & {9.21 $\pm$ 2.18} \\
1323 & dE & 17 & A & - &  / & {2.87 $\pm$ 0.88} & / \\
1369 & dE & 17 &  A & - &  2.83 $\pm$ 1.19 & -0.36 $\pm$ 2.67 & {4.24 $\pm$ 1.17} \\
1488 & E-E/S0 & 17 & S & 7.70 &  3.27 $\pm$ 0.96  & {3.63 $\pm$ 1.0} & {4.75 $\pm$ 1.24} \\
1499 & E-E/S0 & 17 & A & 5.24 & 4.31 $\pm$ 1.08 & {6.01 $\pm$ 1.72} & {6.35 $\pm$ 1.37} \\
1501 & dS0 & 17 & S & {5.68} &  - & - & - \\
1682 & dE & 17 & A & - & / & {3.31 $\pm$ 0.98} & / \\
1794 & dE & 17 & E & - & / & 7.76 $\pm$ 4.90 & {7.55 $\pm$ 1.97} \\
\hline
\end{tabular}
\end{center}
\footnotemark[1]{'-' means that no data were available for this galaxy,
'/' means no line was detected} \\
\footnotemark[2]{Average uncertainties on the $FUV-H$ colours remain below 0.25 mag.} \\
\end{table*}

\section{TTD population: an infrared view}
\label{FIR.sec}
\subsection{FIR detections}

Based on the photometry results from \citet{2013MNRAS.428.1880A} and \citet{2013A&A...552A...8D}, we verify the detection of TTD galaxies by \textit{Herschel}. A galaxy is considered detected in case a 5$\sigma$ flux level is attained in at least one of the five \textit{Herschel} bands. In \citet{2013MNRAS.428.1880A}, flux measurements are determined for the entire VCC catalog covered in the HeViCS observations in an automated way. In a first step, \citet{2013MNRAS.428.1880A} extract subimages of 200 $\times$ 200 pixels centred on the VCC objects. After masking the VCC galaxy and other Virgo objects in the field, 2D polynomials are fit to the masked image to estimate the background level. Polynomials of order 5 are required to reproduce the background in SPIRE images, where foreground cirrus and (un)resolved background sources contribute to the background. A second order polynomial suffices to mimic the background behaviour in PACS maps, which is dominated by instrumental noise. Appropriate ellipses for aperture photometry were derived from an iterative procedure where the edge of the galaxy is defined as the radius outwards of which the signal-to-noise radial profile drops below a value of 2. Fluxes are obtained by measuring the signal within the final ellipses in the background-subtracted maps. Uncertainties on the measured fluxes account for the aperture noise, calibration uncertainty and zero noise. Aperture corrections were applied to both fluxes and uncertainties following the prescriptions in \citet{2010MNRAS.409...38I}. \citet{2013MNRAS.428.1880A} finally perform a visual check by eye to guarantee the precision of the automated measurements. 

\citet{2013A&A...552A...8D} searched the HeViCS maps for far-infrared and submillimeter counterparts of the entire sample of early-type galaxies (ETG) in Virgo. A blind search for detected sources was performed on the SPIRE 250\,$\mu$m image using the source extraction software in HIPE (DAOPhot and SussExtractor). Far-infrared detections were identified by cross-checking the output source catalog with the optical positions of ETGs in Virgo and retaining all extracted sources within 6$\arcsec$ from the {galaxy's optical position} and with a detected flux level of ${\text{S/N}}>5$. In \citet{2013A&A...552A...8D}, flux densities are determined from aperture photometry within 30$\arcsec$ apertures.

Relying on the photometry results from \citet{2013MNRAS.428.1880A} and \citet{2013A&A...552A...8D}, we conclude on the detection of 13 out of 36 TTDs in the HeViCS fields, corresponding to a dust detection rate of 36 $\%$. Twelve out of 13 FIR detections are reported in both papers, {except for VCC\,1684}. The automated photometry pipeline in \citet{2013MNRAS.428.1880A} did not result in the detection of this galaxy, as opposed to \citet{2013A&A...552A...8D} who apply data reduction and photometry techniques specially trimmed for FIR faint objects. All FIR-detections are manifested in the SPIRE 250\,$\mu$m waveband, while the detection rate in other \textit{Herschel} bands is lower (PACS\,100\,$\mu$m: 9/13; PACS\,160$\mu$m: 10/13; SPIRE\,350$\mu$m: 9/13; SPIRE\,500$\mu$m: 6/13). Table \ref{fluxes} summarizes the flux measurements and uncertainties for FIR-detected TTD galaxies, referencing the work for the adopted values. Upper limits correspond to 5$\sigma$ upper limits. 

Owing to the high source densities in the observed regions of the Virgo cluster ($\sim$ 1.11 $\times$ 10$^{-4}$~arcsec$^{-2}$), some of the FIR detections might correspond to the FIR counterpart of background sources rather than TTDs. Statistical tests in \citet{2013A&A...552A...8D} show that about half of the objects detected at flux densities levels $\lesssim$ 64 mJy might be contaminated detections, which applies to six out of thirteen FIR detections. The identification of residual star formation in the selected TTD objects however makes the chance on a possible misinterpretation of their FIR emission unlikely. 

Possible progenitor populations in the VCC catalog with \textit{Herschel} photometry reported in \citet{2013MNRAS.428.1880A} will be used in Section \ref{DustScaling.sec} as comparison sample to analyze the evolution of TTDs based on their dust scaling relations. More specifically, we select objects with late-type galaxy classifications, i.e. blue compact dwarfs (BCD), late-type spirals (Scd-Sd) and Magellanic irregulars (Sm-Im). Galaxies are again considered FIR-detected in case a 5$\sigma$ flux level is attained in at least one of the \textit{Herschel} bands. We refer to \citet{2013MNRAS.428.1880A} for the flux densities of this comparison sample.

\begin{figure*} 
\centering \includegraphics[width=0.455\textwidth]{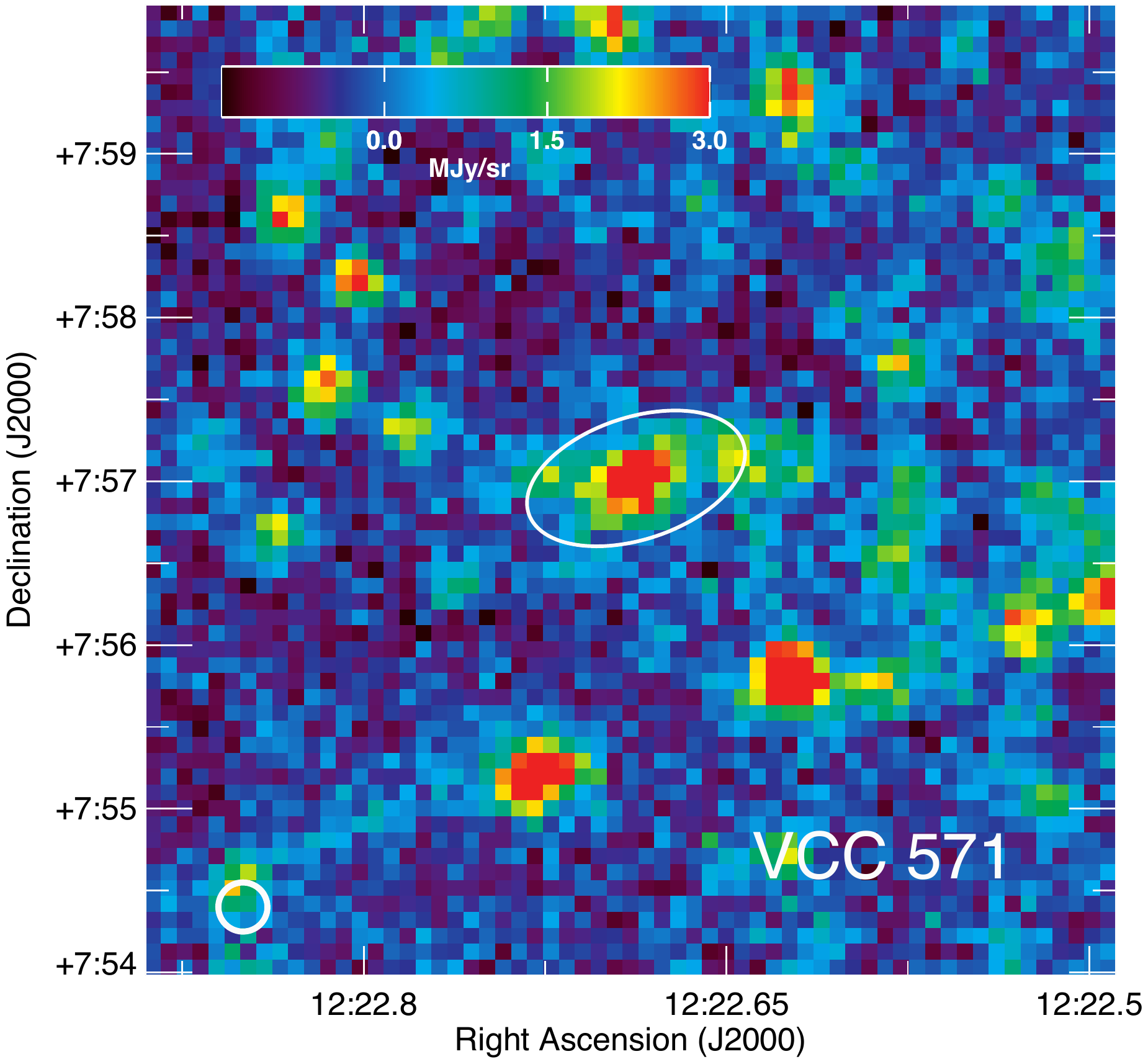}  
\centering \includegraphics[width=0.45\textwidth]{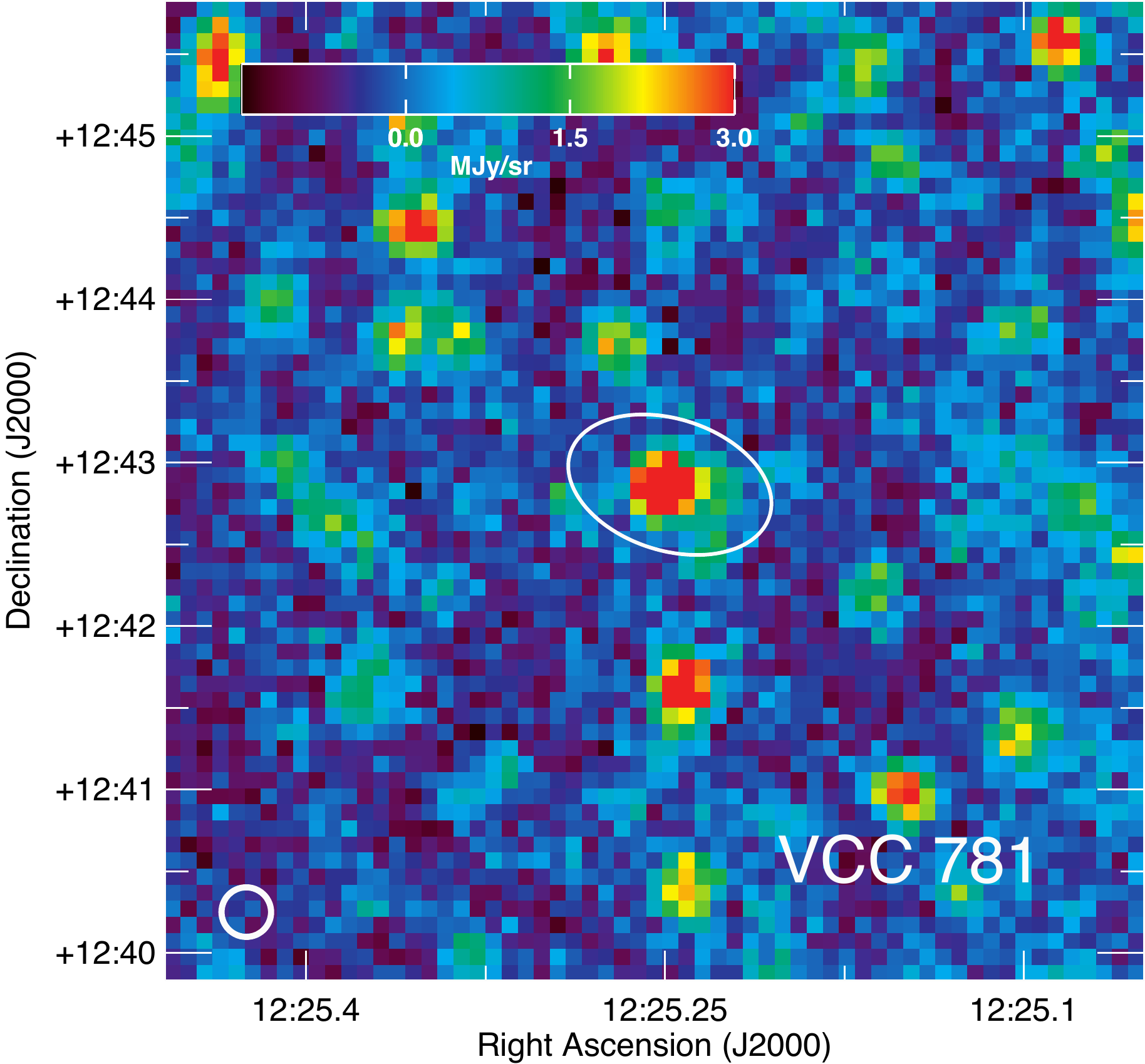}  
\caption{\textit{Herschel} maps of the SPIRE 250\,$\mu$m waveband for transition-type dwarfs, VCC\,571 and VCC\,781. The white ellipses correspond to the optical isophotal diameter $D_{25}$ (25 mag arcsec$^{-2}$), convolved with the SPIRE beam at 250\,$\mu$m. The size of the beam at SPIRE 250\,$\mu$m is indicated in the lower left corner of each panel.}  \label{plot_dust}
\end{figure*}

Figure \ref{plot_dust} shows the \textit{Herschel} detections at 250\,$\mu$m for two TTDs: VCC\,571 (left panel) and VCC\,781 (right panel). The white ellipses correspond to the optical isophotal diameter $D_{25}$ (25 mag arcsec$^{-2}$) as reported by NED. In both galaxies, it is clear that the dust reservoir is confined to the inner regions of the galaxy with the stellar component extending way further out. The dust component is even barely resolved. This suggests that the dust is either compressed in the centres of TTDs during their evolution or removed from the outer regions, which suggests the removal of gas outside-in from the galaxy. The latter scenario seems more plausible in view of the residual star formation confined to the central regions of TTDs (see Section \ref{specphot.sec}). Alternatively, the central star formation might be fueled by a recently accreted gas reservoir, inflowing and flaring up star formation activity in the centres of early-type galaxies (e.g. \citealt{2013ApJ...770L..26D}).

Besides FIR-detected TTDs, eleven low-luminosity early-type galaxies, not identified as being in a transition phase based on their SDSS spectra and with 13.5 $\le$ $m_{\text{pg}}$ $\le$ 18, were detected by \textit{Herschel} at FIR wavelengths \citep{2013MNRAS.428.1880A,2013A&A...552A...8D}. Six of those FIR detections (VCC\,94, VCC\,482, VCC\,672, VCC\,758, VCC\,764, VCC\,1614) have SDSS spectra characterized by a blue continuum and either weak line emission or Balmer absorption, insufficient to pass the selection criteria outlined in Section \ref{Selection.sec}. Either they did not fulfill the selection criteria due to the absence of any recent star formation activity. {Alternatively, any residual star formation might be slightly off-center or rather distributed throughout the galaxy's disk and not confined to their centres, which would fail the selection criteria based on the SDSS spectra taken in the central regions of galaxies with a SDSS spectroscopic fiber of only 3$\arcsec$ in diameter.} Their $NUV-r$ colours (VCC\,94: 4.7, VCC\,482: 5.4, VCC\,672: 4.6, VCC\,758: 5.7, VCC\,764: 4.2, VCC\,1614: 4.6) however suggest they are green valley objects (i.e. objects which are intermediate in color-magnitude space between star-forming galaxies in the blue cloud and quiescent objects on the red sequence), which implies they are physically similar to the TTD population selected in this paper and the origin of dust in these sources is probably similar. The FIR emission from VCC\,292 may be contaminated considering the low surface brightness of the object. The remaining four galaxies (VCC\,270, VCC\,411, VCC\,462, VCC\,486) do not show any clear sign of recent star formation activity. The LINER-like emission lines in VCC\,270 and VCC\,411 suggest that the FIR emission might be biased by the presence of other ionizing sources and not simply originates from dust heated by young stars. The latter four objects are also at the high-mass end of our selection limit with stellar masses of $M_{\star}\sim2-9\times10^{9}~M_{\odot}$ and, therefore, likely not fit into the same evolution scenario proposed for low-luminosity TTDs.

\subsection{Dust masses and temperatures}
\label{modbb.sec}

\citet{2013MNRAS.428.1880A} and \citet{2013A&A...552A...8D} determine dust mass and temperature estimates through the fitting of a modified blackbody function with fixed $\beta$ (2.0), i.e.:
\begin{equation}
\label{modbb}
F_{\nu}~=~\frac{M_{\text{d}}}{D^{2}}\kappa_{\nu}B_{\nu}(T),
\end{equation}
with $F_{\nu}$, the flux density; $M_{\text{d}}$, the dust mass; $\kappa_{\nu}$ = 0.192 m$^{2}$ kg$^{-1}$ \citep{2003ARA&A..41..241D}, the dust absorption coefficient at 350 $\mu$m; $D$, the distance and $B_{\nu}$($T_{\text{d}}$), the Planck function for a dust temperature $T_{\text{d}}$. The $\beta$ value is consistent with the dust emissivity index applied in \citet{2012A&A...540A..52C}. The single-temperature fit for fixed $\beta$ allows a straightforward comparison with the dust masses derived in \citet{2012A&A...540A..52C}, although a dust emissivity index of $\beta$ $\sim$ 1.5 might be more appropriate for star-forming galaxies \citep{2012A&A...540A..54B} and the assumption of a fixed $\beta$ might result in a possible overestimation of the true dust mass. On the other hand, the assumption of a fixed dust emissivity $\beta$ prevents generating spurious results such as the hotspots resulting from resolved SED fits which do not correspond to any heating source \citep{2012MNRAS.425..763G}. The best fitting model with free parameters $M_{\text{d}}$ and $T_{\text{d}}$ is determined from a simplex minimization procedure between the measured fluxes and model fluxes at each wavelength. Model fluxes are determined after applying the appropriate colour corrections factors (see \citealt{2012MNRAS.419.3505D}) and multiplying the modified blackbody functions with the \textit{Herschel} response functions. Only galaxies satisfying the criterium $\chi^{2}_{dof=3}$ $<$ 7.8 were considered as reliable fits. Table \ref{fluxes} includes the resulting dust masses and temperatures from the SED fitting procedures in \citet{2013MNRAS.428.1880A} and \citet{2013A&A...552A...8D}. Dust masses for galaxies with detections for which the SED fitting procedure failed ($\chi^{2}_{dof=3}$ $>$ 7.8) or was not attempted with only three or less data points are determined from Equation \ref{modbb}, the SPIRE 250\,$\mu$m flux density and $T_{\text{d}}$ $\sim$ 18\,K, which approximates the average dust temperature obtained from the SED fits for TTD galaxies reported in \citet{2013MNRAS.428.1880A} and \citet{2013A&A...552A...8D}. We opt for SPIRE 250\,$\mu$m flux densities rather than longer wavelength data (which might be more robust to temperature variations) because the detection of TTDs is not longer guaranteed at longer wavelengths and might be biased by confusion noise. Dust masses determined for this fixed temperature are added to Table \ref{fluxes}, but lack fitting results for the dust temperature. Dust masses and temperatures for the comparison samples of late-type galaxies (BCD, Scd-Sd, Sm-Im) are obtained in the exact same way.

\begin{figure*}
\centering \includegraphics[width=0.46\textwidth]{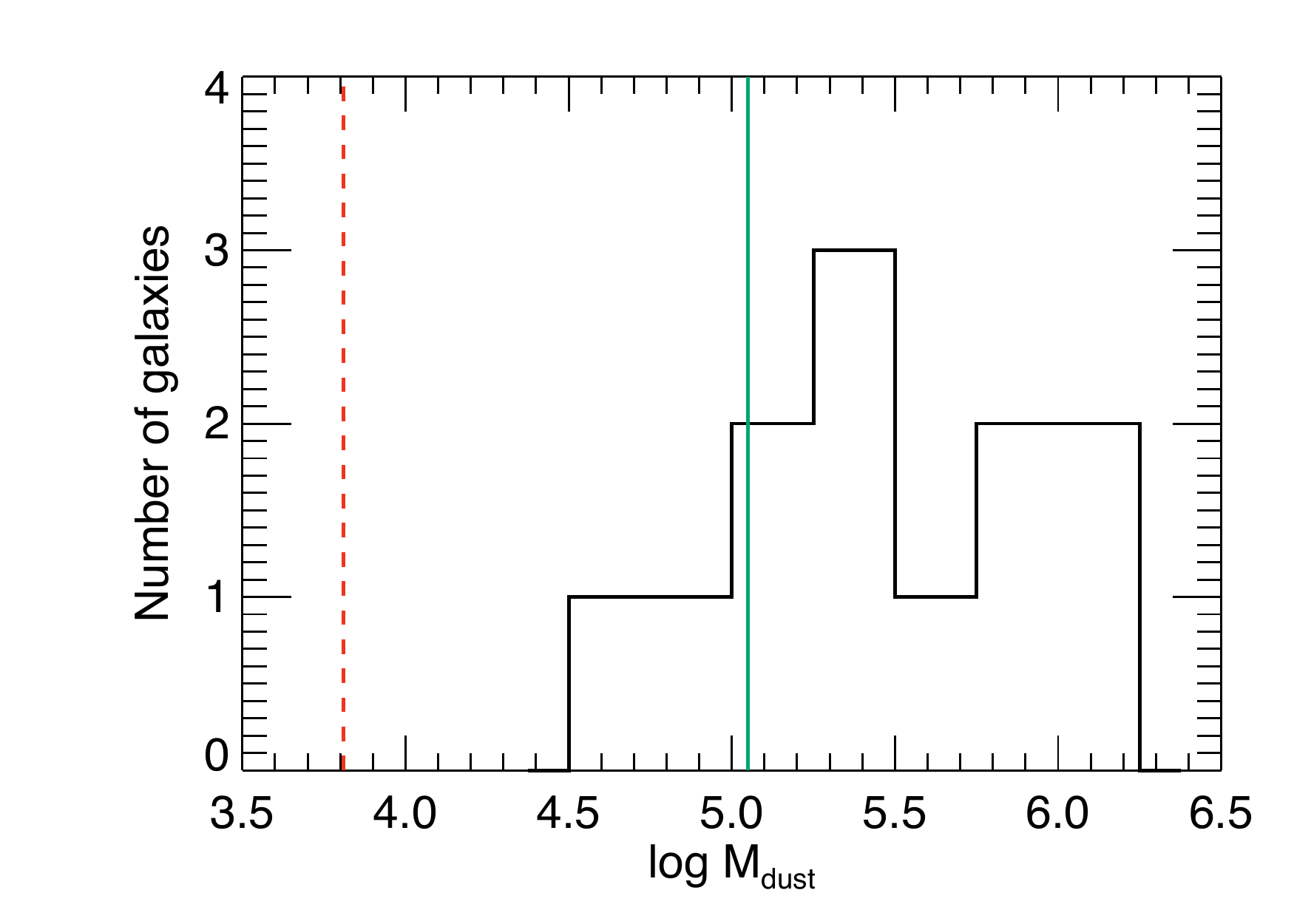}  
\centering \includegraphics[width=0.48\textwidth]{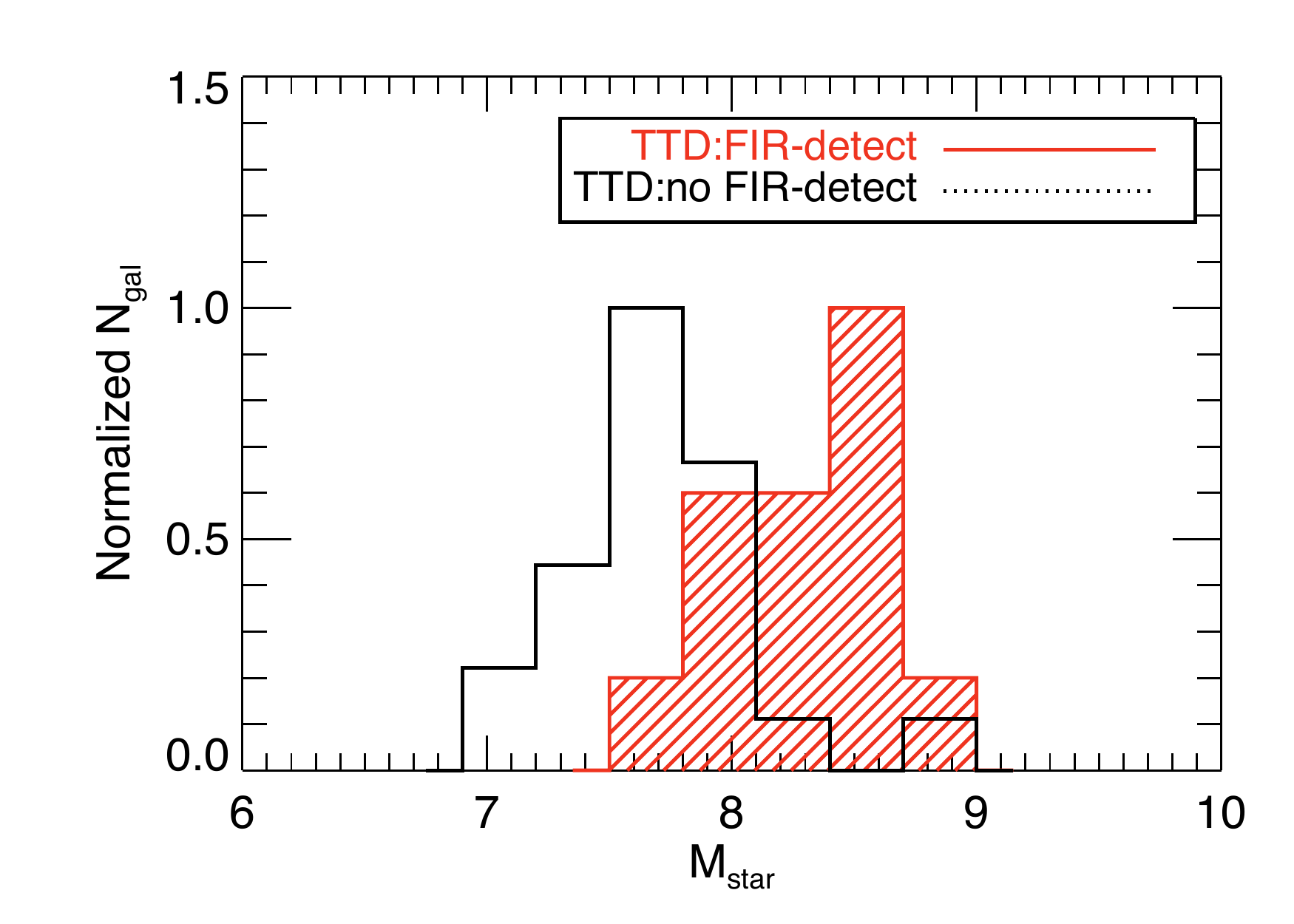}  
\caption{Left: histogram of the dust mass distribution of FIR-detected TTDs. The green solid line represents the 5$\sigma$ dust mass detection limit for the 8-scan HeViCS observations at 250\,$\mu$m, whereas the red, dashed line shows the 5$\sigma$ detection limit obtained from a stacking procedure for quiescent early-type galaxies in the centre of Virgo \citep{2010A&A...518L..54D}. Right: histogram of the stellar masses for FIR-detected (red) and non-detected TTDs (black). Histograms in the right panel are normalized to a maximum of 1 for ease of comparison.} \label{ima_histogram}
 \end{figure*}

Figure \ref{ima_histogram} (left panel) shows the dust mass distribution function for FIR-detected TTDs (black, solid line). The vertical lines indicate the 5$\sigma$ dust mass detection limit in our \textit{Herschel} map at 250\,$\mu$m (green, solid line) and the upper dust mass limit for quiescent early-type dwarf galaxies obtained from a stacking procedure in \citet{2010A&A...518L..54D}. The 5$\sigma$ dust mass detection limit $M_{\text{d}}\sim1.1\times10^5~M_{\odot}$ in the \textit{Herschel} 250 $\mu$m map results from the estimated global noise level of 1$\sigma$ $\sim$ 6.6 mJy/beam in the SPIRE 250 $\mu$m image \citep{2013MNRAS.428.1880A}, assuming a source with temperature $T_{\text{d}}$ $\sim$ 18K at a distance of 17 Mpc. Similarly, the 1$\sigma$ background rms obtained from stacking the FIR maps of early-type dwarfs ($\sim$ 0.38 mJy, \citealt{2010A&A...518L..54D}) is translated into an average 5$\sigma$ upper dust mass limit $M_{\text{d}} \lesssim 6\times10^{3}~M_{\odot}$ for early-type galaxies. Figure \ref{ima_histogram} indicates that FIR-detections are only just above the HeViCS dust mass detection limit, suggesting that a substantial fraction of TTDs might actually harbor dust reservoirs, in contrast with the bulk of genuine ETGs with significantly lower dust masses ($\lesssim 6\times10^3~M_{\odot}$). Some FIR-detected galaxies appear to have a dust content below the \textit{Herschel} detection limit, which is likely an artifact of the uncertainties on the dust mass estimates obtained through SED fitting procedures. The dust mass detection limit is furthermore representative for the average noise level in the SPIRE 250\,$\mu$m. With the noise properties being more favourable in some regions of the map, we might detect objects below this detection limit.

\begin{table*}
\caption{Overview of the photometry and SED fitting results for the FIR-detected TTDs, as reported in \citet{2013MNRAS.428.1880A} and \citet{2013A&A...552A...8D}.
Flux densities correspond to the automated photometry results from \citet{2013MNRAS.428.1880A}. Dust mass and temperature estimates are taken from either \citet{2013MNRAS.428.1880A} or \citet{2013A&A...552A...8D}, in case the former did not quote SED fitting results.}
\label{fluxes}
\begin{center}
\begin{tabular}{@{}|l|c|c|c|c|c|c|c|c|}
\hline 
VCC & $F_{\lambda=100\mu\,m}$ & $F_{\lambda=160\mu\,m}$ & $F_{\lambda=250\mu\,m}$ & $F_{\lambda=350\mu\,m}$ & $F_{\lambda=500\mu\,m}$ & $\log M_{\text{d}}$ & $T_{\text{d}}$ & Ref\footnotemark[1]  \\
 & [Jy] & [Jy] & [Jy] & [Jy] & [Jy] & [$M_{\odot}$] & [K] & \\
\hline \hline 
209 & 0.141 $\pm$ 0.029 & 0.214 $\pm$ 0.033 & 0.233 $\pm$ 0.023 & 0.117 $\pm$ 0.017 & 0.045 $\pm$ 0.010 & 5.9 $\pm$ 0.1 & 16.8 $\pm$ 0.9  & 1 \\
278 & 0.017 $\pm$ 0.004 & $<$ 0.016 & 0.022 $\pm$ 0.004 & $<$ 0.020 & $<$ 0.021 & 5.1\footnotemark[2] & / & 1 \\
327 & 0.120 $\pm$ 0.023 & 0.125 $\pm$ 0.026 & 0.126 $\pm$ 0.021 & 0.060 $\pm$ 0.013 & 0.024 $\pm$ 0.006 & 6.0 $\pm$ 0.2 & 18.8 $\pm$ 1.5  & 1 \\
450 & 0.253 $\pm$ 0.035 & 0.311 $\pm$ 0.047 & 0.219 $\pm$ 0.028 & 0.131 $\pm$ 0.021 & 0.045 $\pm$ 0.010 & 6.0 $\pm$ 0.1  & 19.5 $\pm$ 1.0  & 1 \\
571 & 0.043 $\pm$ 0.013 & 0.107 $\pm$ 0.020 & 0.111 $\pm$ 0.021 & 0.061 $\pm$ 0.016 & 0.019 $\pm$ 0.006 & 5.9 $\pm$ 0.2 & 16.2 $\pm$ 1.0  & 1 \\
710 & 0.101 $\pm$ 0.018 & 0.111 $\pm$ 0.022 & 0.115 $\pm$ 0.015 & 0.087 $\pm$ 0.016 & 0.041 $\pm$ 0.012 & 5.6\footnotemark[2] & / & 1 \\
781 & 0.087 $\pm$ 0.021 & 0.115 $\pm$ 0.022 & 0.078 $\pm$ 0.014 & 0.045 $\pm$ 0.012 & 0.019 $\pm$ 0.005 & 5.4 $\pm$ 0.2  & 19.0 $\pm$ 1.6  & 1 \\ 
788 & $<$ 0.011 & $<$ 0.013 & 0.019 $\pm$ 0.005 & 0.018 $\pm$ 0.005 & $<$ 0.018 & 4.8\footnotemark[2] & / & 1 \\
951 & $<$ 0.015 & 0.081 $\pm$ 0.020 & 0.089 $\pm$ 0.016 & 0.043 $\pm$ 0.011 & $<$ 0.022 & 5.5\footnotemark[2] & / & 1 \\
1175 & 0.016 $\pm$ 0.003 & 0.015 $\pm$ 0.004 & 0.031 $\pm$ 0.009 & $<$ 0.019 & $<$ 0.021 & 5.3\footnotemark[2] & / & 1 \\
1512 & $<$ 0.009 & 0.019 $\pm$ 0.006 & 0.036 $\pm$ 0.008 & $<$ 0.028 & $<$ 0.030 & 5.1\footnotemark[2] & / & 1 \\
1684  & 0.021 $\pm$ 0.016 & 0.041 $\pm$ 0.011 & 0.030 $\pm$ 0.013 & 0.012 $\pm$ 0.010 & $<$ 0.038 & 5.0 $\pm$ 0.3 & 17.8 $\pm$ 2.9  & 2 \\
1715 & $<$ 0.017 & $<$ 0.022 & 0.016 $\pm$ 0.004 & $<$ 0.020 & $<$ 0.018 & 4.7\footnotemark[2] & / & 1 \\
\hline 
\end{tabular}
\end{center}
\footnotemark[1]{References for the dust masses: (1) \citealt{2013MNRAS.428.1880A}; (2) \citealt{2013A&A...552A...8D}}\\
\footnotemark[2]{Dust masses were calculated for a fixed dust temperature $T_{\text{d}}$ $=$ 18K. Accounting for the uncertainty on the assumed dust temperature, we estimate an uncertainty on the dust masses of about a factor of two.}\\
\end{table*}

\subsection{Dust scaling relations}
\label{DustScaling.sec}

\begin{figure} 
\centering \includegraphics[width=0.48\textwidth]{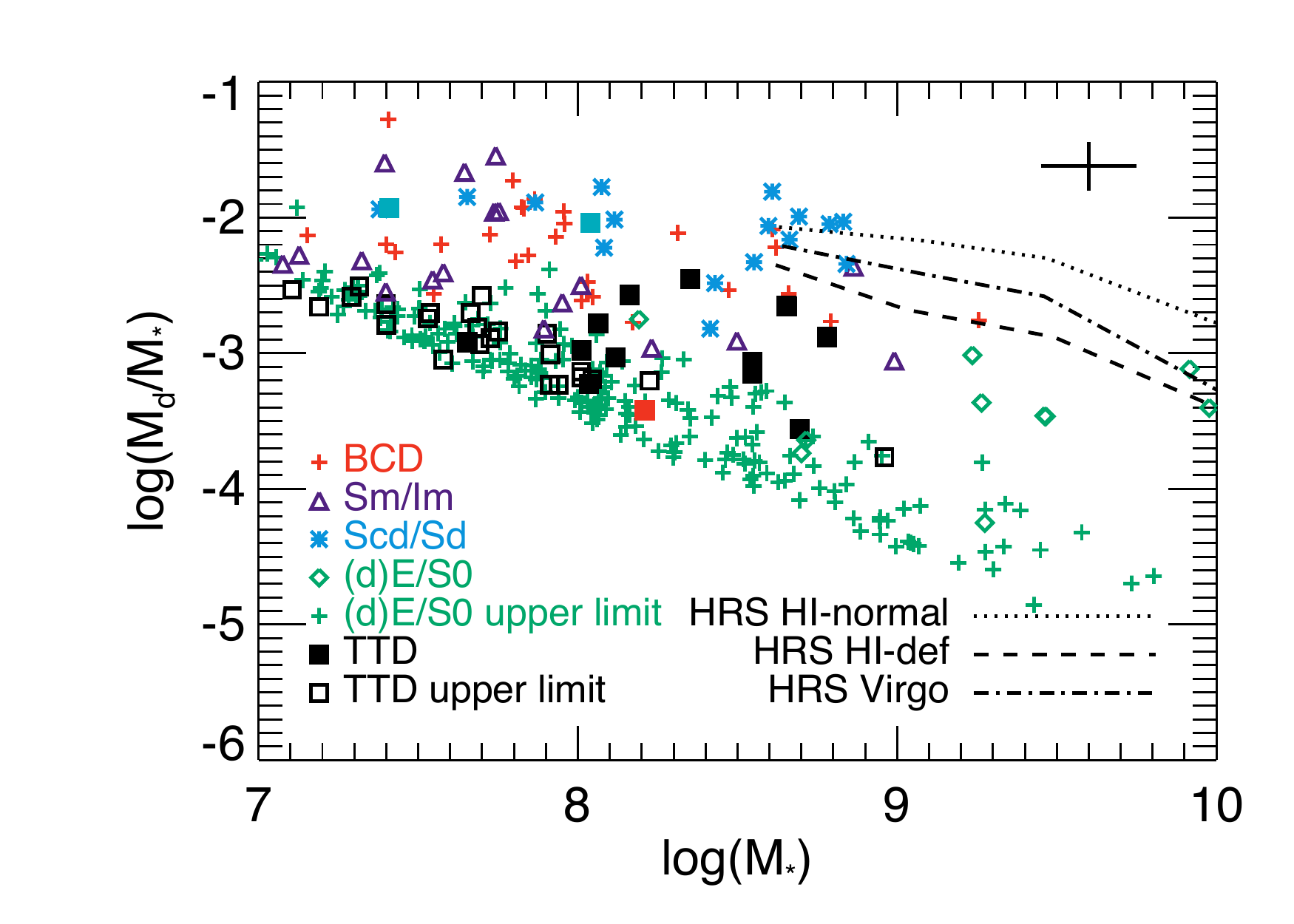}  
\centering \includegraphics[width=0.48\textwidth]{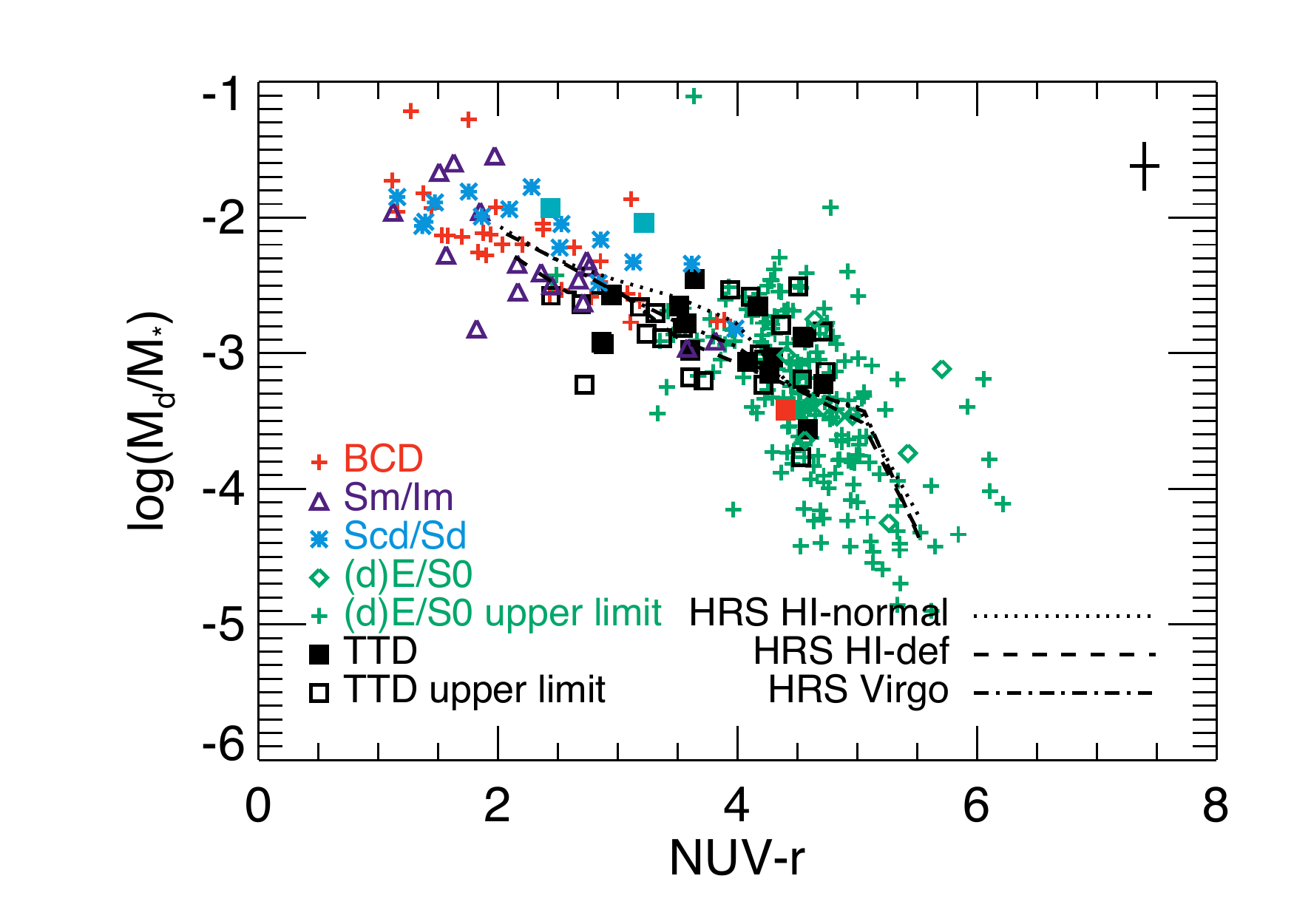}  
\centering \includegraphics[width=0.48\textwidth]{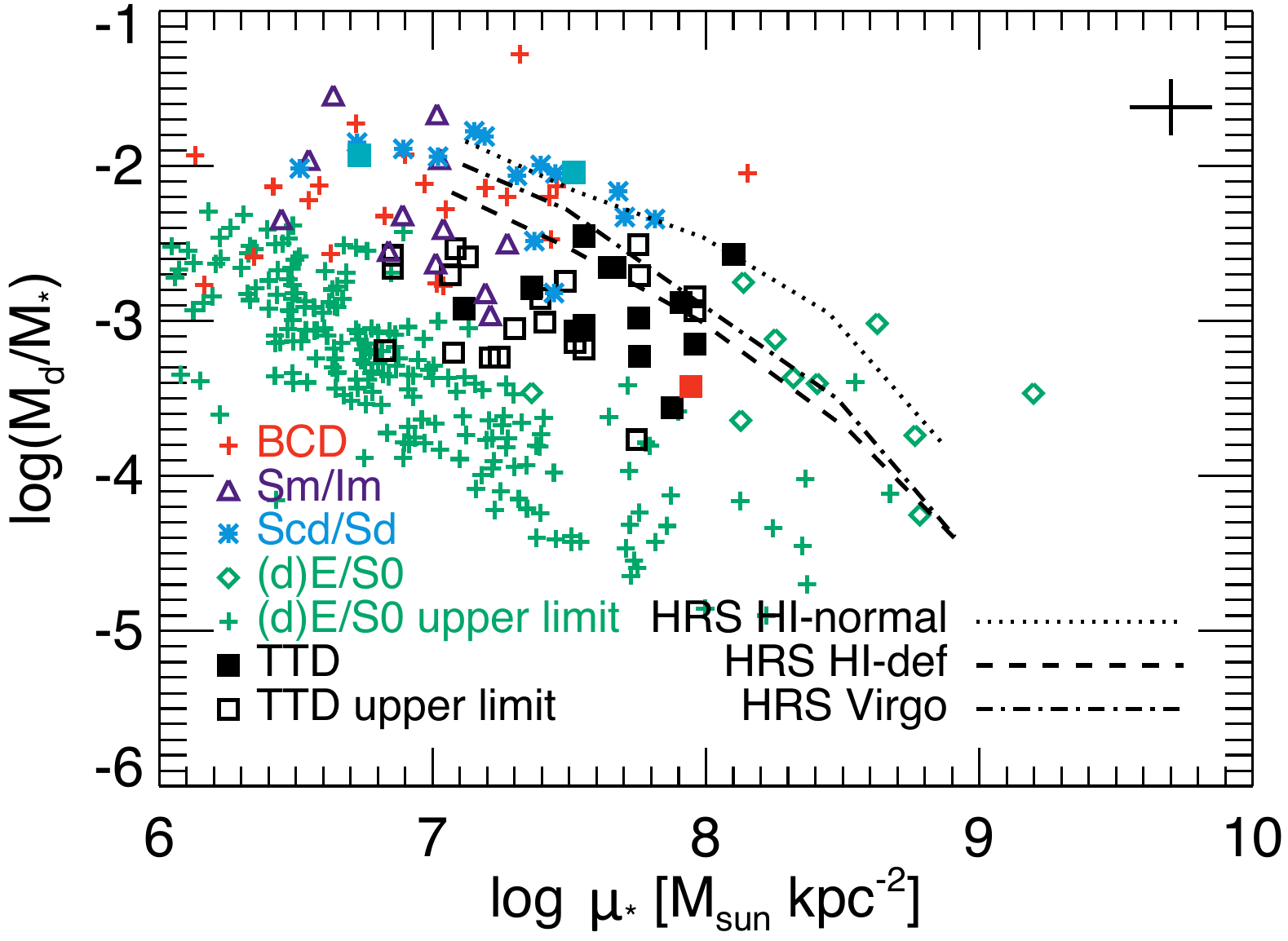}   \\ 
\caption{The dust scaling relations for TTDs, star-forming and early-type galaxies in Virgo. From top to bottom, the panels show the dust-to-stellar mass ratio as a function of stellar mass, $NUV$-$r$ colour (i.e. a proxy for the specific star formation rate) and stellar mass surface density $\mu_{\star}$. }  \label{ima2}
\end{figure}

Dust scaling relations are used to investigate how the dust content in galaxies changes with respect to integrated galaxy properties, such as stellar mass, surface density, metallicity, etc. Such scaling relations can provide insight in the role of dust in the evolution of galaxies and star formation processes throughout the Hubble sequence through a comparison with chemical evolution models and can possibly tell us more about the importance of environmental effects. Dust scaling relations furthermore allow us to investigate the hypothesis that TTDs are formed from infalling late-type galaxies and are gradually evolving into early-type galaxies. In Figure \ref{ima2}, we investigate the scaling of the dust-to-stellar mass ratio as a function of stellar mass (top), $NUV-r$ (middle) and stellar mass surface density $\mu_{\star}$ (bottom) for different galaxy populations. The average observed trends in dust-to-stellar mass ratios for H{\sc{i}}-normal, H{\sc{i}}-deficient and Virgo galaxies from the \textit{Herschel} Reference Survey (HRS, \citealt{2010PASP..122..261B}) are shown as black dotted, dashed and dashed-dotted lines, respectively (see \citealt{2012A&A...540A..52C} for more details).  

Dust scaling relations of the TTD population (filled black squares) are compared to other FIR-detected early-type objects not meeting the TTD selection criteria (green open diamonds) and star-forming late-type galaxies observed by the \textit{Herschel} Virgo Cluster Survey, i.e. blue compact dwarfs (BCD: red crosses), Magellanic irregulars (Sm/Im: purple triangles) and late-type spiral galaxies (Scd/Sd: blue asteriks). Upper limits for undetected TTDs are added to the plots as empty black squares, while undetected early-type galaxies (not classified as TTDs) are indicated as green {crosses}. Only early-type galaxies (FIR-detected and upper limits) with SDSS spectra are shown, to make sure that we are confident that these objects do not pass our selection criteria set for transition-type dwarfs.

Stellar masses are determined from SDSS $i$ band luminosities and mass-to-light ratios, with the latter being derived from the $g$-$i$ colour calibration in \citet{2009MNRAS.400.1181Z}. 
SDSS photometry corresponds to the model fluxes reported on NED. SDSS model magnitudes are derived from fitting a pure exponential profile I(r) = $I_{0}$ exp(-1.68*$r$/$r_{e}$) to the two-dimensional optical data, which should be a good approximation for the nearly exponential light profiles observed in TTDs \citep{2008ApJ...674..742B}. For extended sources, model magnitudes also correspond well to Petrosian magnitudes.
Dust masses are retrieved from \citet{2013MNRAS.428.1880A} and \citet{2013A&A...552A...8D}. Dust mass upper limits are derived from 5$\sigma$ upper limits at 250 $\mu$m, assuming a source at a distance of 17 Mpc and with a dust temperature $T_{\text{d}}$ $\sim$ 18K.
The stellar mass surface density is computed as $M_{\star}$/(2$\pi$ $R_{\text{eff},i}^{2}$) with $R_{\text{eff},i}$ representing the effective radius of the galaxy in the SDSS $i$ band.
Effective radii correspond to SDSS Petrosian radii containing 50$\%$ of the total Petrosian $i$ band magnitude.
Conservative average error bars are indicated in the top right corner of each panel.

The $NUV-r$ colour of galaxies is considered a reliable proxy for the specific star formation rate (sSFR) in late-type systems \citep{2007ApJS..173..315S}.
Correlations between $NUV-r$ and dust fraction were observed for nearby galaxy samples spanning a wide range of different morphological classifications (see also \citealt{2010MNRAS.403.1894D}, \citealt{2012MNRAS.427..703S} and \citealt{2012A&A...540A..52C}) and could be accounted for by appropriate dust formation and evolution models (e.g. \citealt{2008A&A...479..669C}), describing the sensitive balance between the production of new dust grains and its destruction in the interstellar medium. 
With the dust scaling relation in the middle panel of Figure \ref{ima2} being invariant to environmental effects (i.e. H{\sc{i}}-normal and H{\sc{i}}-deficient objects follow the same trend), we are not able to draw any firm conclusions on the deficiency of the ISM in TTDs. 
With $NUV-r$ tracing the variation of star formation (i.e. the evolution of galaxies from the blue cloud to the red sequence along with the diminishing of star formation), the dust scaling relation shows that TTD galaxies on average have $NUV-r$ colours and dust-to-stellar mass ratios in between star-forming galaxies and early-type objects from the HRS sample ($\log M_{\text{d}}/M_{\star} \sim-4.3$, \citealt{2012ApJ...748..123S}), suggesting that TTDs are at an evolution stage intermediate between star-forming late-type and quiescent early-type galaxies. 
{The dust fractions in FIR-detected early-type galaxies ($\log M_{\text{d}}/M_{\star} \sim-3.4$), not classified as TTD based on our selection criteria, are overall consistent with the dust scaling relations for the TTD population ($\log M_{\text{d}}/M_{\star} \sim -2.9$), which again suggests that FIR-detected early-type galaxies constitute a homogeneous class of objects.}

The dust scaling relations with stellar mass and stellar mass surface density (see top and bottom panels in Figure \ref{ima2}, respectively) indicate that the dust fractions are lower in TTDs compared to possible progenitor populations within the same stellar mass and stellar mass surface density range as covered by TTDs. 
Relating this offset of TTDs relative to late-type Virgo objects in the dust scaling relations to the general trends observed for H{\sc{i}}-normal and H{\sc{i}}-deficient HRS galaxies,
we argue that TTDs are relatively gas poor compared to star-forming galaxies which suggests that gas has been stripped from the galaxies or consumed in a very efficient way.
The lower dust fraction indicates that also the dust content is lower, suggesting that the same processes that make TTDs gas deficient also lower their dust masses.
{Alternatively, we can interpret the position of TTDs in the dust scaling relations as an offset from early-type galaxies with the shift of TTDs resulting from recent gas accretion, fueling star formation and making the overall color of TTDs bluer compared to quiescent early-type galaxies.}

Within the same stellar mass range as covered by TTDs ($7.65 \leq \log M_{\star} \leq 8.78$), late-type spiral galaxies ($\log M_{\text{d}}/M_{\star} = -2.11\pm0.28$), BCDs ($\log M_{\text{d}}/M_{\star} = -2.22\pm0.32$) and Magellanic irregulars ($\log M_{\text{d}}/M_{\star} = -2.33\pm0.62$) contain on average more dust than TTDs ($\log M_{\text{d}}/M_{\star} = -2.92\pm0.31$), {requiring the efficient removal} of their dust reservoir to enable a possible link with or transformation into TTDs.
With average dust masses $\log M_{\text{d}}$ of $5.41\pm0.47$ (TTD), $6.24\pm0.42$ (Scd-Sd), $5.85\pm0.35$ (BCD) and $5.57\pm0.39$ (Sm-Im), we would thus require a significant drop in dust mass for star-forming galaxies, in particular for late-type spirals. 
The dust temperature (derived from the SED fitting procedures) does not seem to be significantly different for TTDs compared to star-forming galaxies in Virgo.
We note that this comparison might be biased by the \textit{Herschel} sensitivity limit, allowing us to only compare FIR detections.
The location of the upper limits for undetected TTDs towards the left indeed confirms that the detection of TTDs is biased towards their stellar mass (see top panel in Figure \ref{ima2}). 
Compared to the average dust temperature for elliptical galaxies from the HRS survey ($T_{\text{d}}$ $\sim$ 23.9K, \citealt{2012ApJ...748..123S}), TTDs appear to have colder dust components ($T_{\text{d}}$ $\sim$ 18K). Given that two-thirds of the ellipticals in \citet{2012ApJ...748..123S} show low-ionization narrow-emission line regions (LINERs) and that all of them have hot old stars (post-AGBs and hot-horizontal branch stars), the higher temperatures for HRS ellipticals {not necessarily result} from the presence of young stars.

Performing the same comparison within the stellar mass surface density range covered by FIR-detected TTDs ($7.12 \leq \log \mu_{\star} \leq8.10$), we find dust fractions in late-type spirals ($\log M_{\text{d}}/M_{\star} =-2.22\pm0.29$) and BCDs ($\log M_{\text{d}}/M_{\star}=-2.25\pm0.28$) significantly different from TTDs ($\log M_{\text{d}}/M_{\star} = -2.92\pm0.31$).
Magellanic irregulars have stellar mass surface densities lower than FIR-detected TTDs, which makes it unlikely that they are the immediate progenitors of the FIR-detected TTD population. 
The dust scaling relations however do not rule out that Magellanic irregulars serve as the progenitors of undetected TTDs of lower stellar mass. This result is in agreement with \citet{2012MNRAS.424.2614S}, who argue that the distribution of globular cluster systems (GCSs) in dEs is consistent with the specific frequencies and spatial distribution of GCSs in late-type dwarfs up to stellar masses of $2 \times 10^8~M_{\odot}$.

\section{TTD population: sample characteristics}
\label{Description.sec}
In this section, we analyze the spectrophotometric properties, H{\sc{i}} content, cluster location and heliocentric velocites of the transition-type population in Virgo.

\subsection{Spectrophotometric properties}
\label{specphot.sec}

We rely on a Kolmogorov-Smirnov (K-S) test to verify whether the null hypothesis that two subsamples are drawn from the same parent distribution can be rejected based on their average ages and stellar masses. The spectral line fitting procedure (see \citealt{2007A&A...470..137F}) used for the selection procedure in Section \ref{Selection.sec} provides estimates of the average (luminosity-weighted) stellar age in TTDs. 

At a $>$99$\%$ confidence level, we conclude that the stellar masses are significantly different between the subsamples of FIR-detected and undetected TTDs (see Figure \ref{ima_histogram}, right panel). Assuming that the dust reservoir scales with the stellar mass, the dust content of TTD galaxies is only detected in galaxies with sufficient stellar mass. Based on the average dust-to-stellar mass ratios for FIR-detected TTDs ($\log M_{\text{d}}/M_{\star} = -2.9$), we indeed expect dust masses $M_{\text{d}} \lesssim 10^{5}~M_{\odot}$ for the majority of undetected TTDs populating the stellar mass range $\log M_{\star} \lesssim 8.0$.
The dust detection rate is thus hampered by sensitivity limits, making it hard to draw conclusions on the time scales for evolution in dust masses of TTDs. 
{The K-S test gives a 77$\%$ probability for the null hypothesis that the FIR-detected and undetected galaxies have the same distribution of mean stellar ages, which cannot be interpreted as showing either a similarity or difference in the mean stellar populations between the subsamples.}
{Similarly, a significant difference in $FUV$-$H$ colours (considered to be a diagnostic of the evolution in galaxies throughout the transition-phase, \citealt{2008ApJ...674..742B}) cannot be verified from the K-S test with a probability of 39$\%$ for the null hypothesis that FIR-detected and undetected TTDs are drawn from the same parent distribution.}

The majority of FIR-detected TTDs (8/13)  are classified as galaxies with ongoing star formation, with 3 FIR-detections being in a post-starburst phase. For 2 FIR-detections without SDSS spectra, we could not obtain their spectral classification. Among the undetected TTDs, 15 out of 23 are classified as post-starburst objects and 7 objects show characteristics of ongoing star formation (one object lacks spectral SDSS classification).

\subsection{H{\sc{i}} content}
\label{HI.sec}

The H{\sc{i}} content of TTDs is analyzed by retrieving atomic gas masses or upper limits from the H{\sc{i}} Source Catalog of the Arecibo Legacy Fast ALFA Survey (ALFALFA, \citealt{2011AJ....142..170H}), the Arecibo Galaxy Environments Survey (AGES, \citealt{2012MNRAS.423..787T} and \citealt{2013MNRAS.428..459T}) and the Goldmine database. The ALFALFA and AGES surveys attain average H{\sc{i}} mass detection limits of $\lesssim 3\times10^7~M_{\odot}$ and $\lesssim 8\times10^6~M_{\odot}$, respectively, at the distance of Virgo, 17 Mpc. H{\sc{i}} is detected in only few TTDs by the ALFALFA (VCC\,209: $3.85\times10^{7}~M_{\odot}$; VCC\,304: $3.50\times10^{7}~M_{\odot}$; VCC\,710: $7.78\times10^{7}~M_{\odot}$) and AGES (VCC\,450: $4.57\times10^{7}~M_{\odot}$; VCC\,611: $1.32\times10^{7}~M_{\odot}$) surveys. With H{\sc{i}} masses being close to the H{\sc{i}} mass detection limits of the respective surveys, we might however just be detecting the tip of the iceberg.

For the H{\sc{i}}-detected TTDs, we calculate the H{\sc{i}} deficiency as the logarithmic difference between the expected H{\sc{i}} mass (i.e. ${\text{Def}}_{\text{HI}} = \log M_{\text{HI,ref}} - M_{\text{HI,obs}}$) and the observed H{\sc{i}} content, following the definition in \citet{1984AJ.....89..758H}. We apply calibration coefficients for late-type star-forming objects (Scd-Im-BCD) from \citet{2009A&A...508..201B} relying on their optical diameter to estimate the reference H{\sc{i}} mass, under the assumption that TTDs are the transformation products of star-forming dwarfs. 
A galaxy is considered H{\sc{i}}-deficient in case $Def_{HI}$ $>$ 0.5.
Most of the H{\sc{i}}-detected TTDs seem to be moderately to significantly H{\sc{i}}-deficient (VCC\,209: 1.4; VCC\,304: 1.0; VCC\,450: 1.3; VCC\,611: 1.6; VCC\,710: 0.8). All remaining TTDs are similarly characterized by high H{\sc{i}}-deficiencies ($Def_{\text{HI}}$ $>$ 1.0). Not surprisingly, H{\sc{i}}-detected TTDs all exhibit strong H$\alpha$ emission, characteristic for their ongoing activity of star formation and early stage of TTD evolution. 
While an incompatibility was found between the H{\sc{i}} and quiescent FIR-detected early-type galaxies in \citet{2007A&A...474..851D}, the coexistence of H{\sc{i}} gas and dust in three TTDs suggests a close link between both ISM components, which might imply their shared evolution {(i.e. simultaneous stripping or accretion of gas and dust)}.

The dust-to-H{\sc{i}} ratio inferred for TTDs (VCC\,209: $-1.7$; VCC\,450: $-1.7$; VCC\,710: $-2.3$) is high compared to typical dust-to-gas fractions observed for Magellanic irregular galaxies ($-5 \le \log D/G \le -3$, \citealt{1998ApJ...496..145L}) and blue compact dwarfs ($-4 \le \log D/G \le -3$, \citealt{1998ApJ...496..145L,2002A&A...388..439H}). 
Including submillimeter observations in the SED fitting procedure, \citet{2011A&A...532A..56G} could better probe the total dust reservoir and found $\log D/G$ to vary between $-3$ and $-2$, which is already more consistent with the TTD sample.
The $D$/$G$ estimates for TTDs are considered lower limits due to our assumption on their H{\sc{i}}-dominance and the lack of constraints on their molecular gas content.
With the observed ratios in TTDs being intermediate between the dust-to-H{\sc{i}} fractions observed in H{\sc{i}}-normal (BCD: -2.7, Scd/Sd: -2.0, Sm/Im: -2.2) and H{\sc{i}}-deficient (BCD: -1.7, Scd/Sd: -1.7, Sm/Im: -1.6) galaxies of the HRS sample \citep{2012A&A...540A..52C}, they seem to be in an early evolution stage, having removed only part of their H{\sc{i}} content.

Based on the average $\log D/G \sim -2$ in TTDs and the gas-poor H{\sc{i}} reservoirs ($M_{\text{HI}} < 10^{7}~M_{\odot}$) in undetected TTDs, we expect them to harbour dust reservoirs with masses $M_{\text{d}} < 10^{5}~M_{\odot}$. This upper dust mass limit is consistent with the 5$\sigma$ dust mass detection limit of $M_{\text{d}} \sim 1.1 \times 10^5~M_{\odot}$ for the average noise level of 1$\sigma$ $\sim$ 6.6 mJy/beam at 250 $\mu$m obtained in the four cross-scan \textit{Herschel} observations of the Virgo cluster \citep{2013MNRAS.428.1880A}. This result again shows that the FIR-detection rate of $\sim$ 36$\%$ for TTDs is mainly driven by limitations in sensitivity.

\subsection{Cluster location}

\begin{figure} 
\includegraphics[width=0.47\textwidth]{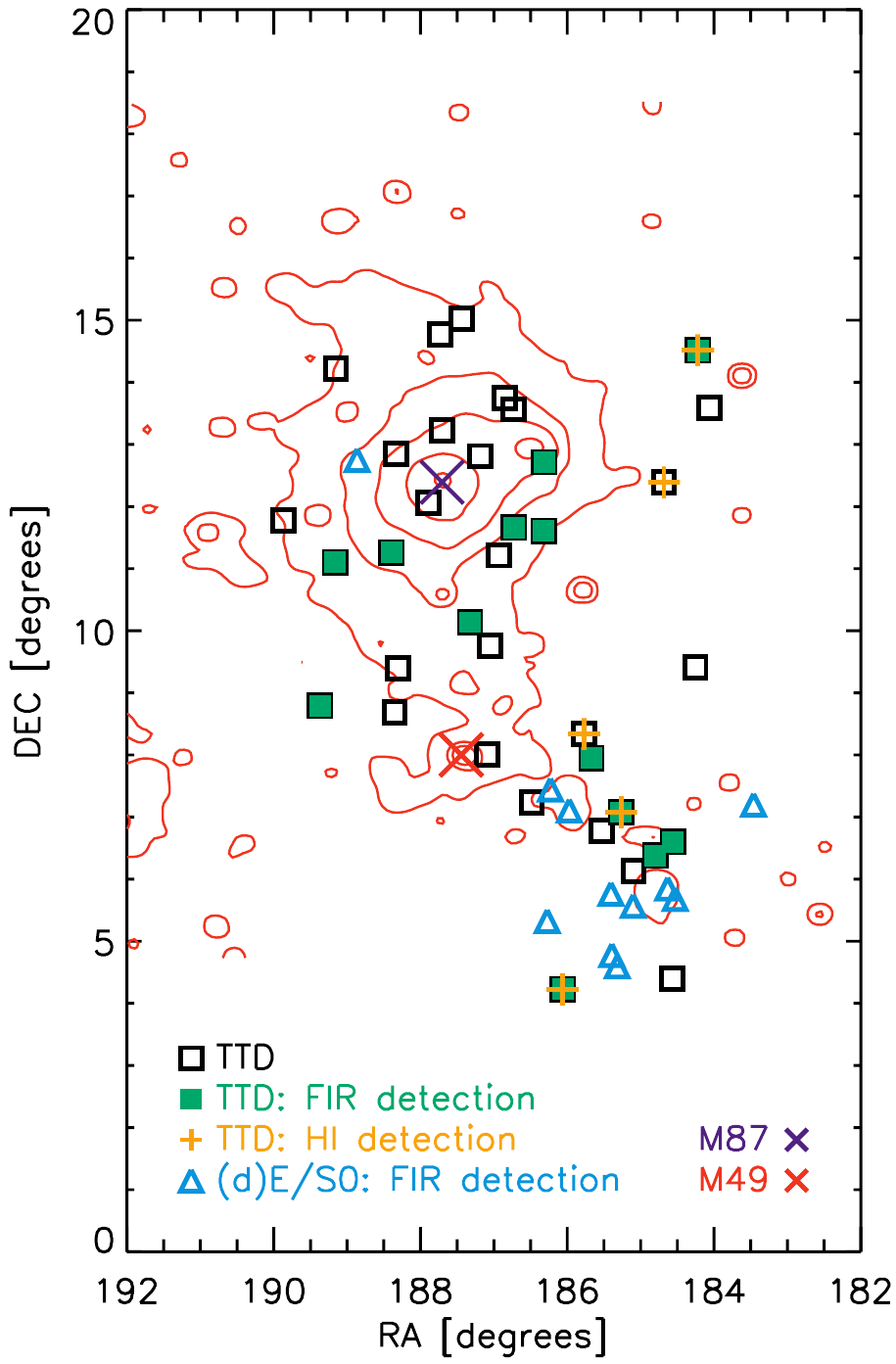}
\caption{Graph of the cluster location for TTDs (black empty squares) in the Virgo cluster. The position of FIR-detected and H{\sc{i}}-detected TTDs are indicated by green filled squares and orange crosses, respectively. Other early-type galaxies (not identified as TTDs) with FIR detections are shown as blue triangles. The X-ray contours from \citet{1994Natur.368..828B} are overlaid as red solid lines on Figure \ref{plot_radec} and indicate the density distribution of hot ionized gas in Virgo.}  \label{plot_radec}
\end{figure}

Figure \ref{plot_radec} shows the location of FIR-detected, undetected and H{\sc{i}}-detected TTDs as well as FIR-detected early-type galaxies (not classified as TTD) within the same photographic magnitude range (13.5 $\le$ $m_{\text{pg}}$ $\le$ 18). Table \ref{transition} indicates the location within subclusters or clouds of TTDs. 
With transition-type galaxies being located in Virgo A and B subclusters as well as in N, W and S clouds, \citet{2008ApJ...674..742B} could show that TTDs resemble the characteristics of a non-virialized population with a degree of clustering that is intermediate between early-type and late-type galaxies.

To verify any significant differences in location between the different subsamples, we rely on a Kolmogorov-Smirnov test comparing the relative distances to M87. With the galaxy density as well as the density of the hot cluster medium decreasing at higher distances from M87, the relative distance to M87 serves as an indicator for the effectiveness and/or incidence of environmental interactions (e.g. ram-pressure stripping, galaxy harassment).
We only perform statistical tests for galaxies observed by \textit{Herschel} and located at a distance of $D$ $=$ 17 Mpc (i.e. excluding galaxies located in the Virgo subcluster B and more distant M and W clouds).
The K-S test gives a 70$\%$ probability for the null hypothesis that the samples of FIR-detected TTDs and non-detected TTDs are characterized by similar relative distances to M87, which does not allow us to distinguish between them based on their cluster location.
{Neither can we differentiate between FIR-detected and H{\sc{i}}-detected TTDs with a probability of 32$\%$ to accept the similarity in their distribution of distances relative to M87. There is a hint for H{\sc{i}}-detected TTDs to be preferentially located at the outer regions compared to non-detections in H{\sc{i}}, with the K-S test giving a probability of 7$\%$ for the null hypothesis claiming similar distributions for both subsamples.}

We furthermore apply statistical tests to verify differences in location between the FIR-detected TTDs and dusty progenitors, i.e. BCDs, late-type spirals and Magellanic irregulars.
{Statistical tests can not rule out that FIR-detected BCDs (52$\%$), late-type spirals (27$\%$) and Magellanic irregulars (79$\%$) are drawn from the same parent distribution as FIR-detected TTDs, with the numbers between parentheses indicating the probability for the null hypothesis to be valid. We, therefore, argue that the cluster location of TTDs and possible progenitors does not contradict the scenario of infalling late-type galaxies that are being transformed by the hostile cluster environment.}

To analyze whether TTDs and progenitors can be distinguished based on their local environment, we compare the number of neighboring galaxies at the same distance within a circular aperture of 1$^{\circ}$ and within 300 km s$^{-1}$ in heliocentric velocity from the galaxy. Local overdensities of neighboring galaxies could be an indication for galaxy harassment to be the key mechanism to transform TTDs. However, the large variety of environments from very dense to poorly populated areas in Virgo for every galaxy population does not allow us to significantly distinguish between their inhabitable regions in Virgo and, thus, does not learn us more about the environmental processes shaping TTDs. 

It thus seems impossible to distinguish between different populations based on their cluster location. 
The latter findings should however be interpreted with caution, since our projected view on the Virgo cluster might bias our analysis. Other than the density from the hot ionized medium, environmental effects such as ram-pressure stripping also depend strongly on the nature of the orbit of galaxies (i.e. galaxies on radial orbits are subject to more efficient stripping, \citealt{2001ApJ...561..708V}).

\subsection{Heliocentric velocities}
\label{Helio.sec}
\begin{figure} 
\includegraphics[width=0.47\textwidth]{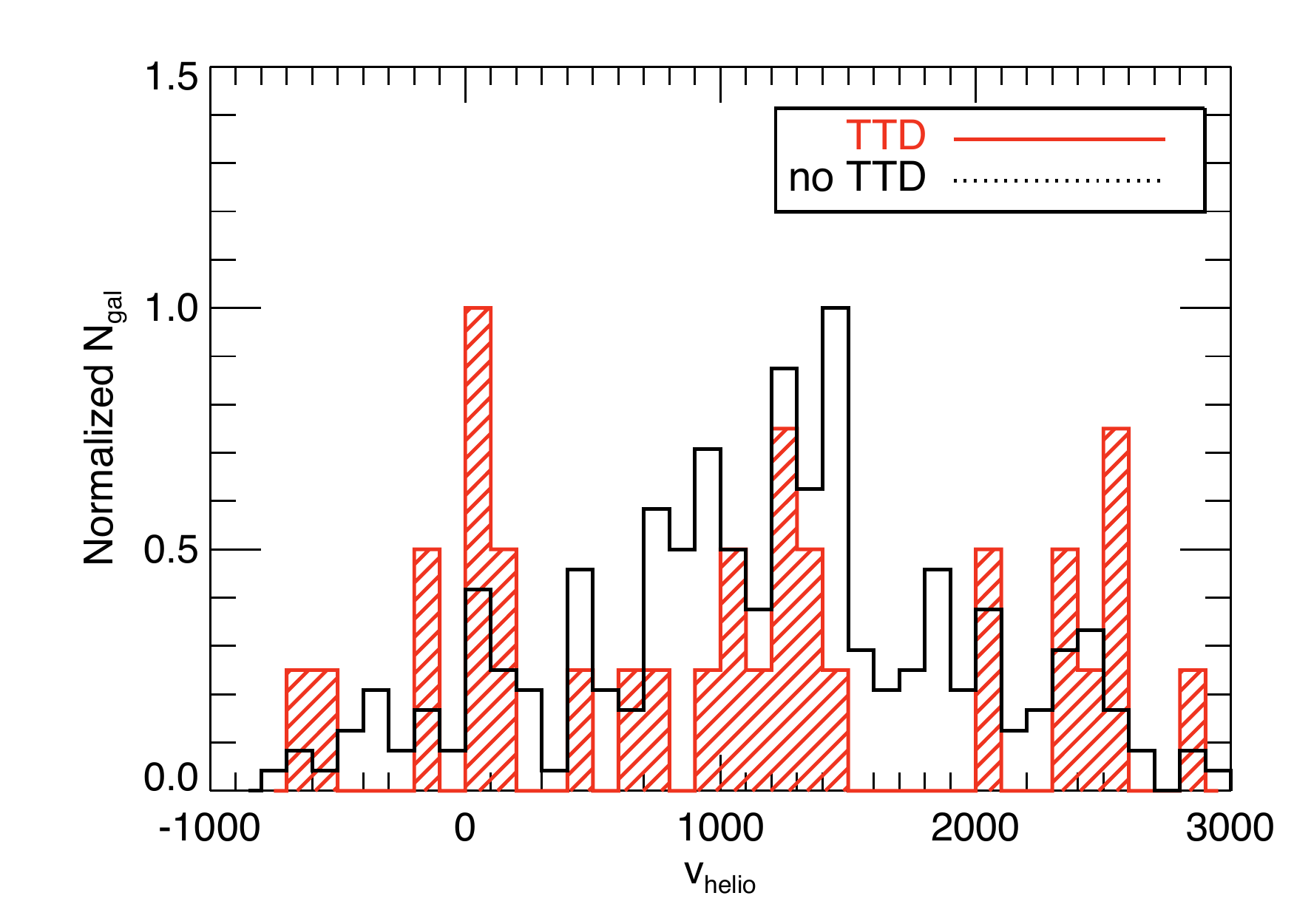}
\caption{Histogram of heliocentric velocities for the TTD sample of galaxies (red filled histogram) and early-type galaxies from our input sample, not classified as TTD (black empty histogram). Histograms are normalized to a maximum of 1 for ease of comparison.}  \label{plot_vhelio}
\end{figure}

Heliocentric velocities allow us to investigate the orbital velocity of objects along the line of sight. Figure \ref{plot_vhelio} represents the histogram of heliocentric velocities for TTDs (red filled histogram) and early-type galaxies from our input sample of 261 low-luminosity objects covered by \textit{Herschel} and with SDSS spectra, but not meeting the selection criteria for TTDs (black empty histogram).
{The velocity distribution of early-types not classified as TTDs has a broad global peak and lower standard deviation as compared to TTDs, which suggests that the former galaxy population is already virialized.   
The velocity distribution of TTDs, on the other hand, has no global peak but is rather characterized by peaks around $\sim$ 0 km s$^{-1}$ and $\sim$ 2500 km s$^{-1}$ suggesting their recent infall onto the cluster.} A similar non-virialized velocity distribution is seen for late-type spiral, Im and BCD galaxies, implying that also the velocity distributions of star-forming galaxies and TTDs are compatible. 

\section{Discussion}
\label{Dustdiscussion.sec}

\subsection{TTDs: transforming late-type galaxies?}
{From an infrared view (see Section \ref{FIR.sec}), the overall picture of the dust scaling relations supports the hypothesis of a transformation between star-forming late-type galaxies to quiescent low-mass ellipticals governed by the cluster environment, coherent with observations in other wavebands. To resemble the dust scaling relations of FIR-detected TTDs, star-forming galaxies need to become H{\sc{i}}-deficient as well as diminish their dust reservoir on short time scales. The removal of gas from galaxies as well as the quenching of star formation is possible on relatively short timescales, but requires efficient mechanisms (e.g. ram-pressure stripping, \citealt{2008ApJ...674..742B,2013arXiv1303.2374G}) to deplete the H{\sc{i}} reservoir (100-300 Myr).}

{Alternatively, the dust scaling relations might rather be telling us that the evolution scenario of TTDs is reversed with galaxies from the red sequence transiting to the green valley due to the recent accretion of gas, which provides a fresh supply for the onset of star formation (e.g. \citep{2012AJ....144...87H,2013ApJ...770L..26D}).
Analyzing the H{\sc{i}}-detected dwarfs in Virgo, \citet{2012AJ....144...87H} can distinguish a blue sample of gas-bearing objects, resembling the properties of star-forming dwarfs, and a red sample, for which the re-accretion of gas seems to be at the origin of their gas content. 
Among the twelve H{\sc{i}}-detected dwarfs in \citet{2012AJ....144...87H}, half of the objects do not have SDSS spectroscopy data and, therefore, could not be identified as possible transition-type objects. The galaxy, VCC\,281, did not enter our initial sample due to its morphological classification of mixed type S/BCD. The SDSS spectrum corresponding to galaxy VCC\,421 has an estimated redshift of $z = 0.4$, which would imply that VCC\,421 is a background source rather than a Virgo cluster member.
Two red sample galaxies (VCC\,1649, VCC\,1993) have SDSS spectroscopy data available, but do not show any sign of star-forming or post-starburst activity and, therefore, do not resemble the characteristics of transition-type dwarfs.
Belonging to the blue subsample, VCC\,304\footnote{Although four TTDs other than VCC\,304 also have a measurable H{\sc{i}} content (see Section \ref{HI.sec}), they did not enter the sample of \citet{2012AJ....144...87H} due to the selection of only sources with $M_{\text{B}}$ $\leq$ -16.} was classified as transition-type object while VCC\,93 only just failed the selection procedure with an H$\beta$ equivalent width of 2.94 $\pm$ 1.09 \AA. While H$\beta$ EW$_{abs}$ $>$ 2.8~\AA\,is sufficient for classification as post-starburst galaxy, the line width measurement did not attain the requested $>$ 3$\sigma$ noise level. Figure \ref{ima2} includes the galaxies from \citet{2012AJ....144...87H} for which we could obtain constraints on their dust content from our \textit{Herschel} observations. Blue gas-bearing dwarfs (VCC\,93, VCC\,281) are indicated as blue filled squares whereas one red galaxy from \citet{2012AJ....144...87H} is presented as a red filled square. While the two blue dwarfs clearly still resemble the properties of late-type dwarfs, red galaxy VCC\,1649 occupies the lower boundary of the range in dust fraction covered by TTDs.
We, therefore, argue that the transition-type galaxies more closely resemble the characteristics of blue, gas-bearing dwarfs in \citet{2012AJ....144...87H}, which are thought to evolve from late-type dwarfs being stripped on their first passage through the cluster. We can, however, not rule out that our sample of TTDs consists of a mixed bag of sources, similar to the H{\sc{i}}-detected early-type dwarfs in Virgo analyzed by \citet{2012AJ....144...87H}.}

The late-type characteristics of TTDs -although classified with early-type morphologies- seem coherent with the scenario of transformation from late-type to early-type objects. Indeed, the majority of TTDs ($\sim$ 56$\%$) harbour structures characteristic of the late-type galaxy population. 
Based on the classifications reported in \citet{2006AJ....132..497L} and \citet{2006AJ....132.2432L} (including 27 out of 36 TTDs from our sample), four objects out of twenty-seven TTDs show evidence for the presence of a disk, while seven other TTDs exhibit substructure other than a disk (see \citealt{2006AJ....132..497L} for more details about the classification). Blue cores were identified in thirteen TTDs based on a visual inspection of their $g$-$i$ colour profiles (see \citealt{2006AJ....132.2432L}).
The identification of structural properties linked to the late-type galaxy population in the TTD sample seem to confirm that the TTD evolution is governed by a transformation from late-type galaxies into quiescent early-type objects.  
Also photometric properties of TTDs support this scenario with $FUV$-$H$ colours being intermediate between the bluer late-type spiral galaxies (purple triangles) and the early-types on the red sequence (red squares for dwarfs and green crosses for more massive objects) (see Figure \ref{plot_UV_H}). This clear distinction in $FUV$-$H$ colour range was exploited by \citet{2008ApJ...674..742B} in their selection procedure for TTDs.

Not surprisingly, all FIR-detected TTDs examined in \citet{2006AJ....132..497L} and \citet{2006AJ....132.2432L} were identified to contain a genuine disk, other substructure and/or harbour a blue central core.
The identification of substructure and/or colour gradients is also reflected in the morphological classification of the transition-type objects. Most of the FIR-detected galaxies are lenticular objects (except for VCC\,788 and VCC\,1715), while the majority of undetected TTDs ($\sim$ 83$\%$) already have the optical appearance of genuine dwarf elliptical galaxies. The morphological classification of FIR-detected TTDs as lens-shaped objects might indicate that part of the low-luminosity lenticular galaxy population represents the intermediate stage of an evolution from late-type to early-type galaxies. 
Also the classification of FIR-detected TTDs with ongoing star formation, while most undetected TTDs are in a post-starburst stage (see Section \ref{specphot.sec}) suggests that star formation has ceased for longer time periods in undetected TTDs or, alternatively, that the star formation simply does not take place in the nucleus of those undetected TTDs. 
The former argument would imply that FIR-detected TTDs are in an earlier phase of evolution, suggesting that a decrease in dust mass occurs simultaneously with the cease of star formation throughout the transition phase. Alternatively, the star formation activity might be more easily discernible in sources of higher stellar masses, implying that the observed trends are merely a sensitivity issue. More sensitive spectral data for the entire galaxy disks of TTDs are required to investigate whether differences in their evolution are significant.  
With also 80$\%$ of the FIR-detected early-type galaxies (not selected as TTDs) being classified as lens-shaped objects rather than ellipticals, we find additional proof to argue that those FIR-detections are not very different from the selected TTD sample.

{The dust scaling relations, blue colors and late-type characteristics are consistent with the radially outside-in quenching of star formation in H{\sc{i}}-rich, star-forming late-type dwarfs that evolve into H{\sc{i}}-poor galaxies with nuclear star formation and finally gas-poor red early-type objects without any star-formation activity (e.g. \citealt{2006ApJ...651..811B,2006AJ....131..716K,2009MNRAS.400.1225C,2012A&A...544A.101C,2013arXiv1303.2374G}). 
Overall the TTD population is characterized by many diverse morphologies, with some of them harboring blue central cores, while other galaxies show disks or other substructures.
The different morphologies suggest that the progenitor population might not be uniform for all TTD galaxies. The progenitor population could either consist of a large variety of late-type galaxies (i.e. late-type spirals, BCDs or dwarf irregulars) which are being stripped on their first passage through the cluster. Alternatively, the optical appearance of TTDs might change throughout the transition phase.}

Ram-pressure stripping has often been put forward as the mechanism governing the evolution and transformation of those H{\sc{i}}-rich late-type galaxies (e.g. \citealt{2008ApJ...674..742B,2013arXiv1303.2374G}). 
In particular, the distinguishable colour gradient characterizing about half of the TTD sample might provide hints for the environmental process at work and/or help to identify the immediate progenitor population. Similar colour gradients were indeed observed in blue compact dwarfs \citep{2006AJ....132.2432L}, which might favour their role as progenitor population. Alternatively, the central blue colours might be a direct consequence of the environmental effects governing the evolution of these objects. Ram-pressure stripping can exhaust the gas component from the outer regions and compress the gas in the inner regions \citep{1972ApJ...176....1G,2009ApJ...694..789T}, giving rise to a central colour excess \citep{2008ApJ...674..742B}.  Also galaxy harassment (e.g. \citealt{1996Natur.379..613M}) is capable of funneling gas to the centre of a galaxy where it forms a density excess, hereby giving rise to the formation of blue central cores \citep{1998ApJ...495..139M,2005MNRAS.364..607M}. Such environmental effects could explain the concentration of the dust reservoir restricted to the centres of TTDs. Starvation, on the other hand, does not seem to provide an evolution scheme consistent with the observed colour gradients \citep{2008ApJ...674..742B}. 

Blue colors and central star formation activity could, however, also emanate from the re-accretion of gas flowing into the center and feeding the centralized star formation \citep{2013ApJ...770L..26D}. In the latter scenario, the dust could be accreted along with the gas and/or produced in the shells of evolved stars and supernovae. Similarly, the dust reservoir of more massive early-type galaxies has argued to be of external origin, acquired through mergers or tidal interactions \citep{1991MNRAS.249..779F,2004ApJS..151..237T,2007ApJ...660.1215T,2012ApJ...748..123S}. Based on the kinematical misalignment of gas and stars in early-type galaxies from the ATLAS 3D Survey \citep{2011MNRAS.413..813C}, \citet{2011MNRAS.417..882D} argued that the external acquisition of gas is important. While mergers (even minor) seem unlikely for the low-luminosity dwarfs in our sample, we can not rule out that some of the TTDs have acquired their cold ISM content through tidal interactions.

\subsection{Transformation process at work}
\label{mech.sec}

{The evolution scenario proposed by \citet{2012AJ....144...87H} for gas-bearing early-type dwarfs in Virgo suggests that gas is accreted through filaments from the intergalactic medium. The accretion of material is considered to occur more efficiently for galaxies moving at low velocity with respect to the cluster at larger distances from the cluster center, since the hot intra-cluster medium in the centre of clusters is expected to prevent efficient accretion processes \citep{2012AJ....144...87H}. Since the TTDs in our sample do not show any excess in gas content (on the contrary, they are all H{\sc{i}} deficient) and most TTDs appear to traverse the cluster on high-speed trajectories (see Section \ref{Helio.sec}), we argue that it is unlikely that an accretion scenario similar to what was proposed by \citep{2012AJ....144...87H} could also apply to the TTD population.
An alternative way of accreting gas and/or dust could take place through tidal interactions with neighboring galaxies. But the high orbital velocity of TTDs in combination with their shallow potential well make it unlikely to attract the ISM material of a companion in a tidal interaction.}

{We, therefore, assume in the remainder of the paper that the observed FIR emission merely originates from cold dust of internal origin and that environmental effects are responsible for the observed properties of TTDs through the transformation of infalling late-type field galaxies. Upon falling onto the cluster, progenitor galaxies are stripped off their outer gas components, which effectively ceases star formation activity and gradually evolves the galaxy from the blue cloud to the red sequence.}
The requirement of efficient removal of H{\sc{i}} gas and quenching of star formation activity on short time scales rules out starvation as dominant transformation mechanism \citep{2008ApJ...674..742B}. Ram-pressure stripping \citep{1972ApJ...176....1G}, viscous stripping \citep{1982MNRAS.198.1007N} and tidal interactions with more massive galaxies and/or the cluster's potential well \citep{2001ApJ...547L.123M,2001ApJ...559..754M,2005MNRAS.364..607M,2006MNRAS.369.1021M} are, on the other hand, capable of governing the observed H{\sc{i}} deficiencies as well as blue central cores. \citet{2008ApJ...674..742B} already designated ram-pressure stripping as favourable transformation mechanism based on their multizone chemospectrophotometric models of galaxy evolution.
Based on the TTD sample selected in this paper, we verify the importance of other mechanisms. 

To infer whether tidal interactions are likely to have influenced TTDs on their orbit through the cluster, we compute the efficiency of this interaction based on the perturbation parameter $P$, which is calculated as
\begin{equation}
\label{eqcluster}
P_{gc}~=~(M_{cluster}/M_{gal})~\times~(R/r_{gal})^{-3}
\end{equation}
for galaxy-cluster interactions ($M_{\text{cluster}}$ is the total cluster mass, $R$ is the distance to the center of the cluster) and
\begin{equation}
P_{gg}~=~(M_{comp}/M_{gal})~\times~(d/r_{gal})^{-3}
\end{equation}
for galaxy-galaxy interactions ($d$ is the intrinsic distance between the interacting galaxies), following the prescriptions reported in \citet{1990ApJ...350...89B}. A perturbation parameter of $P$ $\sim$ 0.006-0.1 is considered critical for the compression of gas towards the center of a galaxy, hereby triggering star formation activity in the central regions of those objects \citep{1990ApJ...350...89B}. We assume a cluster mass of $2.5\times10^{14}~M_{\odot}$ \citep{1999A&A...343..420S} and consider M87 as the center of the cluster. {For a galaxy of mass $10^{10}~M_{\odot}$, the radius ($<$ 4.46 kpc) is estimated from an empirical mass-diameter relation \citep{2006PASP..118..517B}:}
\begin{equation}
\log\left(\frac{M_{\text{gal}}}{M_{\odot}}\right) 
= 
8.46 + 2.37 \log\left(\frac{r_{\text{gal}}}{\text{kpc}}\right).
\end{equation}
Substitution of those values (for $P_{\text{gc}}$ $\sim$ between 0.006 and 0.1) in Eq. \ref{eqcluster} provides an estimate of the influence radius (R $\lesssim$ 1-2.5$^{\circ}$) for tidal interactions with the cluster medium around the central position of M87. 
{Depending on the eccentricity of a galaxy's orbit on its trajectory through the cluster, the velocity and crossing distance from the cluster's centre will differ. Galaxies on radial orbits will traverse the inner regions where the cluster tidal field is strongest and, therefore, perturbations induced by the cluster potential most efficient \citep{2006PASP..118..517B}. Relying on their non-virialized heliocentric velocity distribution (see Section \ref{Helio.sec}), the majority of TTDs is considered to move on elongated, high-velocity radial orbits passing through the central cluster region. Therefore, all transition-type galaxies with a time of residence in the cluster longer than the average cluster crossing time in Virgo ($\sim$ few 10$^{9}$ yr, \citealt{2006PASP..118..517B}) have likely experienced gravitational interactions with the cluster's potential.}

Similarly, we calculate the perturbation parameter for every TTD galaxy and the most massive galaxy in its immediate surroundings. {We consider interactions with neighboring galaxies of $M_{\text{gal}} > 10^{11}~M_{\odot}$, which are on average about 10 times more massive than TTDs. } For each transition-type object, we consider the distance to the closest massive galaxy, which results in perturbation parameters $P_{\text{gg}}$ $<$ 0.001 for most TTDs. {The low efficiencies for galaxy-galaxy interactions suggest that TTDs are currently not under the tidal influence of a nearby, more massive companion.
Given the short duration of tidal encounters in clusters ($\sim$ 10$^{8}$ yr, \citealt{2006PASP..118..517B}), we can not rule out that tidal interactions with more massive cluster members have perturbed the progenitor population of TTDs in the past and are, thus, at the origin of the transformation from low-luminosity late-type systems to early-type dwarfs.}

The stripping of gas via viscous drag (i.e. the cold and dense ISM is {dragged out of the galaxy}, traveling at high speed through the hot intra-cluster medium) could also contribute to the ISM depletion. We argue, however, that the effect of viscous drag is negligible for TTDs, since hydrodynamical simulations have shown that viscous drag becomes important for galaxies in Virgo with linear sizes $r_{\text{gal}}$ $>$ 15 kpc \citep{1984MNRAS.208..261T,1999MNRAS.309..161M} and TTDs have effective radii on average below 1$\arcmin$ (or $\lesssim$ 5 kpc).

In summary, we conclude that a combination of gravitational and hydrodynamical interactions with the cluster medium most likely is responsible for the evolution of the majority of TTDs.
While moving through the cluster's centre, TTDs are affected by the cluster's potential well and the hostile environmental of hot dense X-ray emitting gas. The high velocities of TTDs in infalling clouds can govern a similar evolution in the outer cluster regions with lower gas densities.

\subsection{Evolution in dust fraction of TTDs}
\label{dustev.sec}
{In this section, we assume that TTDs are the descendants of infalling late-type field galaxies of which (part of) the gas reservoir is removed during the first cluster passing, ceasing all star formation activity and evolving to galaxies with an optical early-type morphology.}
Throughout the transition phase, the dust-to-stellar mass ratios in TTDs decreases from values typical for progenitor populations ($\log M_{\text{d}}/M_{\star} \sim -2$) to dust fractions characteristic for quiescent early-type objects ($\log M_{\text{d}}/M_{\star} \lesssim -4$).
Under the assumption of a relatively unchanged stellar mass, the lower dust-to-stellar mass ratio is merely attributed to a reduced dust mass.
This assumption of an unperturbed stellar body is valid for hydrodynamical interactions such as ram-pressure stripping, viscous stripping and starvation. Only gravitational interactions (e.g. tidal interactions, galaxy harassment) might perturb and possible expel part of the stellar content of interacting galaxies. 
Also the stellar mass produced during the transition-type phase ($<$ 1 Gyr, \citealt{2008ApJ...674..742B}) only accounts for a minor fraction of the total stellar mass ($<$ 10$\%$ $M_{\star}$) for typical star formation rates ranging from $10^{-3}$ to $10^{-4}~M_{\odot}$ yr$^{-1}$ (calculated from their $FUV$ luminosities and the SFR calibration in \citealt{Salim}). 
Since the decrease in dust fraction occurs simultaneously with the reddening of TTDs, it seems linked to the quenching of their star formation activity.
{The exact origin of this causality is however not clear, which might be manifested through efficient stripping mechanisms removing the dust simultaneously with the gas. Alternatively, the dust might not be stripped along with the gas, but the destruction of dust and/or reduced grain production will reduce the dust-to-stellar mass ratio after the quenching of star formation.}

{In the latter scenario, we assume that gravitational and/or hydrodynamical interactions have stripped (part of) the gas content of the TTD's progenitor, which eventually will cease all star formation activity. 
While the stripped progenitor galaxy exhausts its remaining gas reservoir through ongoing star formation, stellar feedback through stellar winds and supernovae from massive stars $> 8 M_{\odot}$ might reduce the dust content in TTDs.}
Such energetic feedback processes might destroy the dust due to shocks  or strip the remaining gas/dust from the central regions of the galaxy, where it will be destroyed in the absence of any shielding from the hostile hot interstellar medium (e.g. \citealt{2010A&A...518L..50C}). 
The interstellar medium in TTDs might however consist of cold material (as opposed to the hot ionized gas in early-type systems) which provides a less destructive host for the ejected grains, enabling them to survive for longer time. 

To account for an abrupt drop in dust mass on shorter time scales, the reduction in dust mass needs to occur simultaneously with the cease in star formation activity.
In this case, the dust is stripped simultaneously with the gas from the galaxy through interactions with the hot cluster medium. Environmental effects have shown to be capable of perturbing and possibly expelling part of the dust reservoir from late-type spirals in Virgo \citep{2010A&A...518L..49C,2010A&A...518L..63C,2012A&A...540A..52C,2012A&A...545A..75P}, which make it very plausible for the dust components in lower mass objects to be affected in a similar way. In this scenario, the dust will be able to survive for a longer time since the embedding gas clouds protect the grains from effects capable of destroying them such as radiative or shock sputtering. Once the dust has reached the intra-cluster medium, the survival lifetime of dust is expected to increase due to the lower gas densities \citep{1994Natur.368..828B}. The stripping of gas might however occur more efficient than the removal of dust, due to the more centrally clustered location in the deep potential wells of TTD galaxies, resulting in a left-over dust reservoir in the centres of the TTD galaxies. Rather than being stripped from the galaxy, the left-over dust reservoir will be destroyed due to its exposure to the hostile ISM. 

The amount of dust stripped from the galaxy and, possibly, deposited in the intra-cluster medium depends on the stripping conditions and/or survival lifetime of dust.
The stripping of at least part of the dust components from TTDs seems supported by the concentration of residual star formation and dust in the centres of TTDs as well as the compatibility between H{\sc{i}} and FIR detections.

\subsection{Dust ejection in the intracluster medium}

X-ray observations show that the intra-cluster medium is chemically enriched \citep{1998PASJ...50..187F,2002MNRAS.337...71S,2003ApJ...596..181F,2004ApJ...606L.109F,2004PASJ...56..743H,2004MNRAS.349..952S}. Since the intra-cluster medium is incapable of producing the heavy elements itself, the chemical enrichment seems to originate from cluster galaxies. 
Several mechanisms have been shown to contribute to this transfer from metals in galaxies to their embedding medium both from an observational and theoretical point of view: ram-pressure stripping (e.g. \citealt{2001ApJ...561..708V,2005A&A...433..875R,2006ApJ...651..811B,2006A&A...452..795D,2010A&A...518L..63C}), galaxy-galaxy interaction (e.g. \citealt{2005A&A...438...87K,2008ApJ...687L..69K}), galactic winds (e.g. \citealt{1978ApJ...223...47D,2005AdSpR..36..682K,2006A&A...447..827K,2010A&A...518L..66R}) and intra-cluster supernovae (e.g. \citealt{2004A&A...425L..21D}). \citet{2006A&A...452..795D} showed that ram-pressure stripping processes account for about 10-15$\%$ of the total metal enrichment in the ICM. 
Based on hydrodynamical simulations, \citet{2010MNRAS.409..132W} argue that most of the dust is ejected by lower mass halos with stellar masses $M_{\star} \lesssim 10^{9}~M_{\odot}$, which resembles the stellar mass range covered in our TTD sample. 

In this section, we estimate the total amount of dust ejected by TTDs into the intra-cluster medium (ICM) and analyze to what extent this contributes to the pollution of the ICM. 
The immediate comparison to the observed intra-cluster metals is however hampered by our ignorance on the destruction rate of dust that has been ejected into the cluster medium (see Section \ref{dustev.sec}). The sputtering of dust grains by hot, dense gas might decrease the mass of expelled metals, preferentially destroying smaller dust grains (e.g. \citealt{1991ApJ...381..137F,1994ApJ...430..511S,2012A&A...545A.124B}). Therefore, we consider our estimates as upper limits of the total amount of metals ejected into the IGM.

An estimation of the total expelled dust mass requires a realistic measure of the number of infalling star-forming field galaxies as well as an average mass that is ejected into the ICM by every TTD. 
To estimate the total amount of dust stripped from low-luminosity early-type galaxies since the formation of the cluster,
we rely on the slope of the optical luminosity function ($\alpha$ = -1.40, \citealt{1985AJ.....90.1759S}) for low-luminosity ($B_{T}$ $\geq$ 14) early-type galaxies in the Virgo cluster (see their Figure 13). 
Here, we safely assume that all low-mass early-type galaxies originate from the transformation of star-forming late-type galaxies, which have fallen onto the cluster since its formation. 
The current ejection rate of dust into the ICM is predicted from the present-day infall rate of $\sim$ 300-400 galaxies per Gyr (e.g. \citealt{2008ApJ...674..742B} and \citealt{2013arXiv1303.2846G}).
Accounting for the infalling galaxies in the past ($\sim$1150 galaxies) and at the present epoch ($\sim$350 galaxies), we can estimate the total amount of dust expelled into the intra-cluster medium by low-luminosity early-type dwarf galaxies. 

Assuming a Schechter function can describe the stellar mass distribution of low-luminosity galaxies in the field in a similar way as the luminosity function ($\alpha$ varies between -1.30 and -1.50, \citealt{2005ApJ...631..208B}), we can estimate the dust mass in unperturbed progenitors by assuming a reasonable constant dust fraction $\log~M_{\text{d}}/M_{\star} \sim -2$. Based on the high end of the stellar mass range of progenitors in Virgo ($< 10^{9}~M_{\odot}$) and sampling the incomplete faint end of the galaxy population ($m_{pg}$ $\geq$18), we assume a stellar mass range $10^{5}~M_\odot \leq M_{\star} \leq 10^9~M_{\odot}$ for infalling field galaxies, which roughly corresponds to dust masses between $10^{3}~M_{\odot}$ and $10^{7}~M_{\odot}$.
For a slope $\alpha$ = -1.40, cut-off  dust mass $M_{\text{d}} = 10^{7}~M_{\odot}$ and number density of 10 at the knee of the Schechter function, we can calculate the number of galaxies in each of the dust mass bins $M_{\text{d}} = 10^{3}~M_{\odot}$ (917), $10^{4}~M_{\odot}$ (365), $10^{5}~M_{\odot}$ (144), $10^{6}~M_{\odot}$ (52) and $10^{7}~M_{\odot}$ (10). 
Under the assumption that the entire ISM content of progenitors is stripped, we can estimate the total amount of metals ejected by infalling field galaxies ($\sim 1.7\times10^{8}~M_{\odot}$). {Due to the unvirialized nature of their velocity distribution, most late-type galaxies and spirals are considered to have fallen onto the cluster during the last few Gyr \citep{1987AJ.....94..251B}. Based on semi-analytic models, \citet{2012MNRAS.423.1277D} however argue that those galaxies could reside in clusters already for 5 to 7 Gyr.
Assuming galaxies have fallen onto the cluster during the last 3 to 7 Gyr, this translates into average dust ejection rates of 0.06 and 0.02 $M_{\odot}$ yr$^{-1}$, respectively.
We furthermore argue that this dust ejection rate is an upper limit since some of the infalling field galaxies belong to compact groups where efficient pre-processing (e.g. tidal interactions between the group members) might have taken place and expelled part of the ISM content of some group members.}

Our estimated upper limit is somewhat lower compared to the estimated amount of dust grains ejected by individual intergalactic stars ($\sim 0.17~M_{\odot}$ yr$^{-1}$) {and similar to the predicted dust contribution from more massive early-type galaxies in clusters ($0.04~M_{\odot}$ yr$^{-1}$), following the predictions from \citet{2000A&A...354..480P}. }
\citet{2006A&A...452..795D} estimate mass loss rates based on a combination of N-body and hydrodynamical simulations for the cold ISM disks of galaxies, depending on the cluster mass and its merger history. Combined mass loss rates for all galaxies in a cluster seem to vary from $> 1~M_{\odot}$ yr$^{-1}$ ($1.3\times10^{15}~M_{\odot}$) to $\sim 0.2~M_{\odot}$ yr$^{-1}$ ($8.7\times10^{14}~M_{\odot}$), depending on the cluster mass assumed in their simulations. Knowing that the cluster mass of Virgo ($2.5\times10^{14}~M_{\odot}$, \citealt{1999A&A...343..420S}) is even smaller, the theoretical mass loss rates are considered to be even lower than 0.2 $M_{\odot}$ yr$^{-1}$, suggesting that the low-luminosity galaxies stripped by the cluster medium contribute significantly to the enrichment of the cluster in the central 2.5 Mpc compared to other galaxy populations. The numerical predictions in \citet{2006A&A...452..795D} furthermore show that ram-pressure stripping can account for most of the metal enrichment in the central 100 kpc of clusters, which would imply that infalling low-luminosity objects influenced by the dense cluster medium constitute the primary source for metal enrichment in the central regions of clusters.

\section{Conclusions}
\label{Conclusions.sec}

In this paper, we select 36 transition-type dwarf galaxies (TTDs) in the Virgo cluster based on SDSS spectral diagnostics characteristic for their ongoing/residual star formation.
We study the far-infrared properties of the TTD population based on \textit{Herschel} observations as part of the \textit{Herschel} Virgo Cluster Survey.
Thirteen TTDs are detected by \textit{Herschel}, translating into a dust detection rate of $\sim$ 36$\%$.
With FIR-detections populating the high end of the stellar mass range covered by TTDs and assuming that the dust mass scales with the stellar mass, this detection rate is limited by the 5$\sigma$ dust mass detection limit $< 1.1 \times 10^{5}~M_{\odot}$ in our HeViCS observations, suggesting that also undetected TTDs harbor relatively massive dust reservoirs in contrast to the average upper dust mass limit in genuine early-type galaxies ($M_{\text{d}} < 6 \times 10^{3}~M_{\odot}$). 

Dust scaling relations indicate that the evolution of dust properties with respect to stellar mass, surface density and $NUV-r$ colour (i.e. a proxy for the specific star formation rate, sSFR, and reddening) is intermediate between late-type and early-type galaxies, confirming the hypothesis of infalling field galaxies transforming to quiescent low-mass elliptical objects governed by environmental effects. 
Blue compact dwarfs, Magellanic irregulars and late-type spirals could act as the progenitors of the TTD population based on their observed FIR properties.
TTDs demonstrate blue central cores as well as central dust concentrations, suggesting a radially outside-in removal of gas and quenching of star formation activity by environmental processes such as ram-pressure stripping and galaxy harassment.
The scaling relations furthermore show that the evolution in dust emission is essentially related to the evolutionary sequence in star formation. Our FIR observations however do not allow to determine the origin for this correlation, i.e. star formation declines as a consequence of ISM removal or the lack of a dusty ISM results from the cease of star formation activity. 

{We can, however, not rule out that the dust scaling relations rather indicate a reversed evolution scenario with galaxies from the red sequence transiting to the green valley due to the recent accretion of gas, which provides a fresh gas supply for the onset of star formation in the centres of TTDs.}
 
{In the scenario of TTDs descending from infalling late-type field galaxies through environmental effects,} the decrease in dust fraction throughout the transition phase suggests the destruction and/or stripping of a significant part of the dust reservoir.
In the first scenario, some dust is not removed along with the H{\sc{i}} gas that is stripped from the galaxy, after which it quickly will be destroyed in sputtering processes. 
In case of dust removal along with the stripped gas, metals are ejected into the intra-cluster medium, where they are capable of surviving in a lower-density environment, shielded from the hostile environments in embedding gas clouds. 
{In case dust is being stripped along with the gas, we predict about $1.7 \times 10^{8}~M_{\odot}$ of metals to be ejected into the ICM throughout the lifetime of the cluster or an equivalent dust ejection rate of $\sim 0.02-0.06~M_{\odot}$ yr$^{-1}$ over the last 3 to 7 Gyr, suggesting the importance of low-luminosity systems contributing to the ICM pollution in the centres of clusters.}

\section*{Acknowledgements}    
{We thank the referee for his/her comments on the paper, which
helped to improve the paper.}
IDL is a postdoctoral researcher of the FWO-Vlaanderen (Belgium).
MG gratefully acknowledges support from the Science and Technology Foundation (FCT, Portugal) through the research grant PTDC/CTE-AST/111140/2009.
DP acknowledges the kind hospitality at the MPE.
PACS has been developed by a consortium of institutes
led by MPE (Germany) and including UVIE
(Austria); KU Leuven, CSL, IMEC (Belgium);
CEA, LAM (France); MPIA (Germany); INAFIFSI/
OAA/OAP/OAT, LENS, SISSA (Italy);
IAC (Spain). This development has been supported
by the funding agencies BMVIT (Austria),
ESA-PRODEX (Belgium), CEA/CNES (France),
DLR (Germany), ASI/INAF (Italy), and CICYT/
MCYT (Spain). SPIRE has been developed
by a consortium of institutes led by Cardiff
University (UK) and including Univ. Lethbridge
(Canada); NAOC (China); CEA, LAM
(France); IFSI, Univ. Padua (Italy); IAC (Spain);
Stockholm Observatory (Sweden); Imperial College
London, RAL, UCL-MSSL, UKATC, Univ.
Sussex (UK); and Caltech, JPL, NHSC, Univ.
Colorado (USA). This development has been
supported by national funding agencies: CSA
(Canada); NAOC (China); CEA, CNES, CNRS
(France); ASI (Italy); MCINN (Spain); SNSB
(Sweden); STFC and UKSA (UK); and NASA
(USA).

\bsp  

\label{lastpage}

\end{document}